\documentclass[aps,twocolumn, amsmath,amssymb,pra,superscriptaddress]{revtex4-2}

\usepackage{graphicx}
\usepackage{dcolumn}
\usepackage[dvipsnames]{xcolor}
\usepackage{bm}

\usepackage{empheq,etoolbox}
\usepackage{bbold}
\usepackage{appendix}

\usepackage[section]{placeins}

\DeclareFontFamily{U}{mathb}{\hyphenchar\font45}
\DeclareFontShape{U}{mathb}{m}{n}{
      <5> <6> <7> <8> <9> <10> gen * mathb
      <10.95> mathb10 <12> <14.4> <17.28> <20.74> <24.88> mathb12
      }{}
\DeclareSymbolFont{mathb}{U}{mathb}{m}{n}

\DeclareMathSymbol{\Earth}{3}{mathb}{"43}

\newcommand{\p}{\partial}
\renewcommand{\vec}{\mathbf}

\usepackage{verbatim}
\usepackage{xcolor}
\definecolor{lightgrey}{rgb}{0.8, 0.8, 0.8}
\definecolor{misprint}{rgb}{0, 0, 1}
\definecolor{remark}{rgb}{0.0, 0.5, 0.69}

\newcommand{\changesOK}[1]{\textcolor{black}{#1}}

\begin{document}

\title{\changesOK{Analytical description of}\\ collisional decoherence in a BEC double-well accelerometer}
\author{Kateryna Korshynska}
\affiliation{Institut für Mathematische Physik, Technische Universität Braunschweig, Mendelssohnstraße 3, 38106 Braunschweig, Germany}
\affiliation{Fundamentale Physik für Metrologie FPM, Physikalisch-Technische Bundesanstalt PTB, Bundesallee 100, 38116 Braunschweig, Germany}
\affiliation{Department of Physics, Taras Shevchenko National University of Kyiv, 
64/13, Volodymyrska Street, Kyiv 01601, Ukraine}
\author{Sebastian Ulbricht}
\affiliation{Fundamentale Physik für Metrologie FPM, Physikalisch-Technische Bundesanstalt PTB, Bundesallee 100, 38116 Braunschweig, Germany}
\affiliation{Institut für Mathematische Physik, Technische Universität Braunschweig, Mendelssohnstraße 3, 38106 Braunschweig, Germany}

\begin{abstract}


BEC-based quantum sensors offer a huge, yet not fully explored potential in gravimetry and accelerometry. In this paper, we study a possible setup for such a device, which is a weakly interacting Bose gas trapped in a double-well potential.
In such a trap, the gas is known to exhibit Josephson oscillations, which rely on the coherence between the potential wells. 
Applying the density matrix approach, we consider transitions between the coherent, partially incoherent, and fully incoherent states of the Bose gas. 
\changesOK{
We provide an analytical description of the collisional decoherence due to weak interactions, causing the Josephson oscillations to decay with time.
In particular, we give the mathematical link between
that decay in the density matrix approach and its interpretation in terms of phase fluctuations.
To investigate the potential of the double-well setup as a quantum sensor we apply additional external acceleration to the system.
The interplay of collisional interaction and acceleration leads to an additional shift of the oscillation frequency. 
We give the analytical expression for this shift and estimate the sensitivity of a hypothetical BEC double-well accelerometer based on that effect.
}
\end{abstract}

\maketitle

\section{Introduction}

Modern cooling and trapping techniques allow for a precise manipulation of cold boson gases \cite{phillips1991optical}. When the ultra-cold temperatures are reached, a Bose gas forms a macroscopic quantum system known as a Bose-Einstein condensate (BEC). After its first realization \cite{anderson1995observation, davis1995bose}, the study of BECs has become a rapidly evolving field of research, providing new opportunities for both fundamental science and technological applications.

One of the most attractive features of a BEC is its long-range coherence, leading to pronounced interference phenomena. This feature makes BECs a promising basis for quantum sensor devices with huge, yet only partially explored, potential in gravimetry and accelerometry \cite{gaaloul2014precision,PhysRevA.111.043308,chaika2025acceleration}. 
One realization of such a device is a BEC confined in a double-well potential.
The tunneling dynamics of particles in this potential leads to Josephson oscillations between the wells, which are highly sensitive to even small accelerations of the trap \cite{masi2021multimode, gersemann2020differential, PhysRevLett.117.203003}. 

\changesOK{Commonly, a Bose gas of many interacting particles, is efficiently described by a single, fully coherent quantum state, which obeys the Gross-Pitaevskii equation \cite{pitaevskii1961vortex, gross1961structure} and constitutes the ground state of the system.
In addition, up-to-date techniques such as ZNG formalism \cite{griffin2009bose} and the stochastic Gross-Pitaevskii equation \cite{gardiner2002stochastic,Bidasyuk}
are capable to describe a small fraction of the Bose gas, forming a thermal cloud of excited particles.
These approaches are well suited, as long as the
trapped Bose gas has well-separated energy levels, and most of the bosons occupy the ground state, as it is the case in a single-well, e.g. in a harmonic potential.
However, an interferometer setup such as a double-well has
two or more nearly degenerate energy levels, which are macroscopically occupied.
In this case, due to decoherence, the state of system does not stay in a 
coherent superposition of the energy eigenstates \cite{sakmann2014universality, fattori2008atom}.
The theoretical description of this decoherence poses a challenge to the commonly used approaches to BEC dynamics.}

The aforementioned problem has been previously discussed in the non-interacting limit for a Bose gas confined in a double-well trap \cite{PhysRevA.109.043321}. We have shown that in the density matrix formalism, the Josephson equations are modified by the presence of partial incoherence, leading to the observable modifications to Josephson dynamics. In the present paper, we extend this research to the case of a closed, weakly interacting Bose gas, investigating how the collisional interaction induces decoherence in the system. \changesOK{The collisional decoherence in a Josephson junction has been previously investigated numerically in Refs.~\cite{PhysRevA.55.4318, PhysRevLett.87.180402},
using a phase fluctuations approach.
While in the coherent regime the phase between the potential wells is clearly determined,
it fluctuates in consequence of the decoherence  \cite{PhysRevLett.87.180402, pitaevskii2016bose,castin1997relative, RevModPhys.90.035005}. We relate these phase fluctuations to the density matrix approach, elucidating the connection between the two formalisms.}

The paper is organized as follows. In Sec.~II we introduce a double-well geometry and its single-particle eigenstates and analyze how they are modified by the presence of external acceleration. Moreover, from these eigensates we construct the left and right well states, which provide a suitable basis to describe the Josephson effect. Then, in Sec.~III we write down the many-particle Hamiltonian of the Bose gas, which consists of the free single-particle Hamiltonians and collisional interactions, which we treat as a small perturbation. With that in hand, we solve the Liouville equation for the density matrix in Sec.~IV and derive the effective density matrix that contains all information relevant to study Josephson effect. 
In Sec.~V we 
investigate the Josephson dynamics 
for the particular scenario, where all particles are initially localized in the left well, and demonstrate the collisional decoherence. 
Moreover,  
we study the influence of external acceleration, causing observable frequency shifts,
and discuss the setup's
suitability as an accelerometer device. 
In Sec.~\ref{Sec: Comparison with the standard two state model} we compare the obtained frequency shifts with the predictions of the pure state model, which is applicable in case of negligible decoherence. \changesOK{In Sec.~\ref{Sec:Density matrix and phase fluctuations approach}  we analyze the correspondence between the density matrix and the phase fluctuations approach.}
The conclusions are made in Sec.~\ref{Sec: Conclusions}.

\section{A single particle in a double-well}
\label{Sec: A single particle in a double-well}

Since we want to study a Bose gas in the weakly interacting regime, it is convenient to first investigate the single-particle solution to the double-well problem, build up from that the many-particle theory and treat interactions perturbatively, later. 
In this Section, therefore, we introduce the trap geometry, and give the single-particle energy spectrum as well as the corresponding eigenstates. 

\subsection{Two-state model for 3D cubic traps}
\label{Sec: Two-state model for 3D cubic traps}

\begin{figure}[htp]
    \centering  \includegraphics[scale = 0.4]{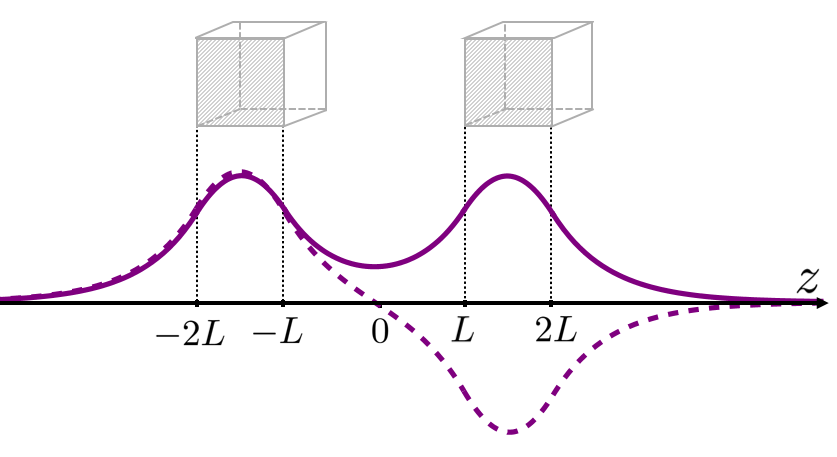}
    \caption{Potential geometry and the single particle energy eigenstates: the ground state $\phi_0$ (solid line) and the excited state $\phi_1$ (dashed line).}
    \label{Fig: trap geometry}
\end{figure}

While our theory can be applied to any double-well potential shape,
in our model we consider two cubic traps, each of volume $L^3$ and potential depth $U_0$, as well as a spacing of $2L$ between them, as shown in Fig.~\ref{Fig: trap geometry}. This geometry is inspired by the recent studies of BECs confined in box-shaped potentials, see e.g., Refs.~\cite{navon2021quantum, PhysRevLett.110.200406}, and is used here to illustrate our ideas by means of an instructive  example.

For simplicity, we want the potential depth $U_0$ to be tuned such that the trap configuration has only two bound states $\phi_0 (\mathbf{r})$ and $\phi_1 (\mathbf{r})$, solving the eigenvalue equation $\hat h_0(\vec{r})\phi_{0/1}=E_{0/1}\phi_{0/1}$ for the Hamiltonian $\hat h_0(\vec{r})=\vec{p}^2/2m+U(\vec{r})$.
The details on parameters and solutions for this system can be found in Appendix~\ref{Appendix: Potential geometry and single-particle states}.
For our further discussion, it is relevant that these states are symmetric and asymmetric along $z$ direction, respectively, as shown in Fig.~\ref{Fig: trap geometry}. Moreover, their energies $E_0$ and $E_1$ are nearly degenerate, such that $\Delta E = E_1-E_0\ll |E|$, where $E=(E_0+E_1)/2$.
For completeness, the extension of the formalism beyond the two-state model can be found in  Appendix~\ref{Appendix: Beyond two state model}.

\subsection{Two-state model in the presence of external acceleration}
\label{subsec: Two-state model in the presence of external acceleration}

In the following, we assume that the double-well is subject to an external acceleration $\vec{a}$, creating the additional potential $\delta U(\mathbf{r}) = m \vec{a}\cdot \vec{r} =ma (z \cos \theta + x \sin \theta)$, where $\theta$ is the angle between $\vec{a}$ and the $z$-axis.

Applying perturbation theory \cite{binney2013physics} for the nearly degenerate states $\phi_{0/1}$, we find that
\begin{subequations}
\label{eq: perturbed states}
\begin{eqnarray}
    \psi_0 &=& \sqrt{\frac{\Delta E^a + \Delta E}{2\Delta E^a}} \phi_0 - \textrm{sign}_a \sqrt{\frac{\Delta E^a - \Delta E}{2 \Delta E^a}} \phi_1 \label{eq: perturbed states 0}\\
    \psi_1 &=& \sqrt{\frac{\Delta E^a + \Delta E}{2\Delta E^a}} \phi_1 + \textrm{sign}_a \sqrt{\frac{\Delta E^a - \Delta E}{2\Delta E^a}} \phi_0 \label{eq: perturbed states 1}\,,
\end{eqnarray}
\end{subequations}
obey the eigenvalue equation $[\hat h_0(\vec{r})+\delta U(\vec{r})]\psi_{0/1}=E^a_{0/1}\psi_{0/1}$ to all orders in $maL/\Delta E$ and to linear orders in $maL/E$ and $\Delta E/E$. In Eqs.~(\ref{eq: perturbed states}) we introduced
$\Delta E^a = \sqrt{(\Delta E)^2 + (2ma\chi(\theta))^2}$, $\chi(\theta) = \left[\int d\mathbf{r} \phi_0 \phi_1 z\right] \cos \theta$, and $\textrm{sign}_a = \textrm{sign}[a \chi (\theta)]$. 
The corresponding energies are
\begin{equation}
    E_{0/1}^a = \frac{1}{2} \left(E_0 + E_1 + 2ma I(\theta) \mp \Delta E^a \right)\, ,
    \label{eq: asymmetric enery levels}
\end{equation}
where $I(\theta) = \left[\int d\mathbf{r} \phi_0^2 x\right] \sin \theta$. 

The obtained result has the same structure as the solution of the slightly asymmetric double-well problem, given in  Ref.~\cite{PhysRevA.109.043321}, where
$2ma\chi(\theta)$ plays role of the potential step.
In this scenario, however, the latter is caused by the acceleration component parallel to the $z$-axis, while the orthogonal acceleration component only causes a global energy shift $maI(\theta)$ in Eq.~(\ref{eq: asymmetric enery levels}).

In most of all cases, where the two boxes are well separated (such that $\phi_0$ and $|\phi_1|$  almost resemble each other),
we can approximate $\int d\mathbf{r} \phi_0^2 x\approx \int d\mathbf{r} \phi_0 \phi_1 z\approx 3L/2$, i.e.,  $\chi(\theta) \approx 3L \cos \theta/2$ and $I(\theta) \approx 3L \sin \theta/2$. 
Note that the perturbative approach, leading to Eqs.~(\ref{eq: perturbed states}) only holds for small accelerations, i.e. for
$ma\chi (\theta)\approx 3maL \cos \theta/2 \ll \Delta E$. For larger accelerations, the eigenstates of the Hamiltonian $\hat{h}_0 + \delta U$
cannot be properly approximated as superpositions of $\phi_{0/1}$ states.

\subsection{Left-right states}

In this Section, we introduce the left-right state basis $\psi_{L/R}$ of the single-particle states (in the case $a=0$ denoted as $\phi_{L/R}$) that maximize the probability of the particle to be in the left or right well, respectively. Following \cite{PhysRevA.109.043321}, they read $\psi_L=\phi_L = (\phi_0 + \phi_1)/\sqrt{2}$ and $\psi_R=\phi_R = (\phi_0 - \phi_1)/\sqrt{2}$ in the basis of the unperturbed energy eigenstates $\phi_{0/1}$.
However, in the presence of acceleration, we need to express them in the basis of the proper energy eigenstates $\psi_{0/1}$, what gives rise to 
\begin{subequations}
\label{eqn:psi_LR}
    \begin{eqnarray}
    \psi_L &=& \frac{1}{2\sqrt{\Delta E^a}} \left([\sqrt{\Delta E^a + \Delta E} - \textrm{sign}_a  \sqrt{\Delta E^a - \Delta E}]\psi_0 \right.\nonumber \\ 
     &+& \left. [\sqrt{\Delta E^a + \Delta E} + \textrm{sign}_a \sqrt{\Delta E^a - \Delta E}]\psi_1\right), \\
    \psi_R &=& \frac{1}{2\sqrt{\Delta E^a}}\left([\sqrt{\Delta E^a + \Delta E} + \textrm{sign }_a \sqrt{\Delta E^a - \Delta E}]\psi_0 \right. \nonumber\\ 
     &-& \left. [\sqrt{\Delta E^a + \Delta E} - \textrm{sign}_a \sqrt{\Delta E^a - \Delta E}]\psi_1\right).
\end{eqnarray}
\end{subequations}
For the sake of brevity, we 
introduce the angle $\xi$, determined by  $\tan \xi = (\Delta E^a + \textrm{sign}_a \sqrt{(\Delta E^a)^2 - (\Delta E)^2})/\Delta E$, to write
\begin{equation}
        \begin{bmatrix} 
	|\psi_{L} \rangle  \\
        |\psi_{R} \rangle \\ 
	\end{bmatrix} =
       \begin{bmatrix} 
	\cos \xi |\psi_{0}\rangle + \sin \xi |\psi_{1} \rangle  \\
    \sin \xi   | \psi_{0} \rangle - \cos \xi |\psi_{1} \rangle \\ 
	\end{bmatrix} = \hat{T}(\xi)\begin{bmatrix} 
	|\psi_{0}\rangle   \\
        | \psi_{1} \rangle \\ 
	\end{bmatrix}
 \label{eq:left-right states}\,.
\end{equation}
The transformation matrix $\hat{T}$ will be of later use, when we discuss the Josephson dynamics in the presence of external acceleration.

\section{Many-particle Hamiltonian}
\label{Sec: Perturbation theory}

Having discussed a single particle in a double-box trap, in the following we build up a description of a closed many-particle system, to first consider a non-interacting Bose gas.
Afterwards, we treat additional interparticle interactions perturbatively, applying the effective density matrix formalism.

\subsection{Non-interacting Bose gas}
\label{Subsec: Non-interacting Bose gas}

The Hamiltonian $\hat{\mathcal{H}} = \hat{\mathcal{H}}_0 + \hat{\mathcal{V}}$ is dominated by the non-interacting part
\begin{equation}
    \hat{\mathcal{H}}_0 = \sum_{i=0}^N [\hat{h}_0 (\mathbf{r}_i)+\delta U(\mathbf{r}_i)]= \sum_{i=0}^1 E^a_i \hat{a}_i^\dagger \hat{a}_i,
    \label{Eq: unperturbed many particle hamiltonian}
\end{equation}
where $\hat{a}^\dagger_i$ and $\hat{a}_i$ are creation and annihilation operators associated with the two energy eigenstates ($i = 0,1$) of the double-well. The interaction $\hat{\mathcal{V}}$ is a small perturbation due to the collisional interaction of the particles that will be discussed in Sec.~\ref{Subsec: Collisional interaction}.

Now we introduce the density matrix for the non-interacting many-particle system. For that, we need to construct the basis of many-particle states
\begin{eqnarray}
    |N_1\rangle &=& \frac{1}{\sqrt{N_1! (N-N_1)!}} \left(\hat{a}_0^\dagger\right)^{N-N_1} \left(\hat{a}_1^\dagger\right)^{N_1} |0\rangle\,, \label{eqn:many-particle_eigen_states}
\end{eqnarray}
which are labeled by the number $N_1$ of particles in the excited state $\psi_1$. The states 
satisfy $\hat{\mathcal{H}}_0|N_1\rangle=[N_1 E^a_1+(N-N_1)E^a_0]|N_1\rangle=:\mathcal{E}_{N_1}|N_1\rangle$, and
are normalized such that $\langle N_1'|N_1\rangle=\delta_{N_1'N_1}$.
With these states at hand, we can construct the density matrix 
\begin{eqnarray}
    \hat{\rho} &=& \sum_{N_1,N_1'} \rho_{N_1N_1'}|N_1 \rangle \langle N_1'| \label{eqn:density_matrix}
\end{eqnarray}
that contains the full information about the $N$-particle system.

\subsection{Collisional interaction}
\label{Subsec: Collisional interaction}
To describe the interparticle interactions, we need to find the corresponding interaction potential, entering the Hamiltonian $\hat{\mathcal{H}} = \hat{\mathcal{H}}_0 + \hat{\mathcal{V}}$. 

Commonly, BECs are dilute gases, where the average distances between particles are large in comparison to the range of interparticle interactions. 
Thus, in such a gas one can consider only 2-particle interactions, while simultaneous interactions involving more particles are very rare and can be neglected \cite{pitaevskii2016bose}.
This also means that the interaction is significant only if the particles closely approach each other, as can be modeled by a delta-potential 
\begin{equation}
    V(\mathbf{r}_1 - \mathbf{r}_2) = g \delta(\mathbf{r}_1 - \mathbf{r}_2) \,. \label{eqn:Vofr}
\end{equation}
This assumption is also supported by the theory for elastic scattering of slow particles (i.e., low temperature) \cite{pitaevskii2016bose, griffin2009bose}, where the interaction constant $g = 4\pi \hbar^2 a_s/m$ is found to be determined by the scattering length $a_s$. 
The latter is a crucial parameter, defining the interaction properties of a Bose gas, and can be accessed with a few experimental techniques, see e.g. \cite{tiesinga1996spectroscopic}.
\begin{figure}[b!]
    \centering  \includegraphics[scale = 0.5]{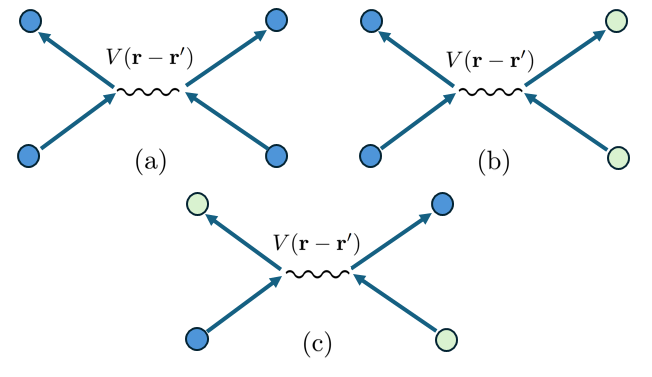}
    \caption{Collisional processes in Bose gas: (a) collisions of type $\hat{a}_i^\dagger \hat{a}_i^\dagger \hat{a}_i \hat{a}_i$, where both ingoing and outgoing bosons belong to the same energy eigenstate $\epsilon_i$ (depicted by same color); (b) Hartree (direct) collisions and (c) Fock (exchange) collisions $\hat{a}_i^\dagger \hat{a}_j^\dagger \hat{a}_i \hat{a}_j$ of bosons in different states (depicted by different colors).}
    \label{Fig: Collision diagrams}
\end{figure}

The two-particle collisions described by the potential (\ref{eqn:Vofr}) enter our formalism via the operator
\begin{equation}
    \hat{\mathcal{V}} = \frac{1}{2}\sum_{i j r l}V_{ijrl} \hat{a}_i^\dagger \hat{a}_j^\dagger \hat{a}_r \hat{a}_l.
    \label{eq: all 2 body interactions}
\end{equation}
with the matrix element
\begin{eqnarray}
    V_{ijrl} &=& \int d \mathbf{r}_1 d \mathbf{r}_2 \psi^\star_i(\mathbf{r}_1) \psi^\star_j(\mathbf{r}_2) V(\mathbf{r}_1 - \mathbf{r}_2) \psi_r (\mathbf{r}_2) \psi_l (\mathbf{r}_1) \nonumber \\ &=& g \int d \mathbf{r} \psi^\star_i(\mathbf{r}) \psi^\star_j(\mathbf{r}) \psi_r (\mathbf{r}) \psi_l (\mathbf{r})\, , \label{eqn:matrix_element}
\end{eqnarray}
which characterizes a collision between two ingoing bosons in the states $\psi_{i/j}$ that are found in the states $\psi_{r/l}$ after interaction \cite{morgan2000gapless}.
Here the $\psi_i(\vec{r})$ with ($i = 0,1$) are given by the wave functions for non-interacting particles from Eq.~(\ref{eq: perturbed states}),
as can be done in the considered approximation of weak interactions
\cite{mazets2010thermalization}. 

Since we consider the Bose gas as a closed system with fixed total energy and particle number, we will only consider 
such collisional processes, which are
energy-conserving, leaving us with the interaction operator
\begin{eqnarray}
    \hat{\mathcal{V}} &=& \frac{1}{2} \left(V_{0000} \hat{a}_0^\dagger \hat{a}_0^\dagger \hat{a}_0 \hat{a}_0 + V_{1111} \hat{a}_1^\dagger \hat{a}_1^\dagger \hat{a}_1 \hat{a}_1 \right.\nonumber \\ 
     &+& \left. V_{1010} \hat{a}_0^\dagger \hat{a}_1^\dagger \hat{a}_0 \hat{a}_1 +  V_{0101} \hat{a}_0^\dagger \hat{a}_1^\dagger \hat{a}_0 \hat{a}_1 \right. \label{eqn:interaction_operator}\\ 
     &+& \left. V_{1001} \hat{a}_0^\dagger \hat{a}_1^\dagger \hat{a}_0 \hat{a}_1 +  V_{0110} \hat{a}_0^\dagger \hat{a}_1^\dagger \hat{a}_0 \hat{a}_1 \right) .\nonumber
\end{eqnarray}
The corresponding collisional processes are illustrated in Fig.~\ref{Fig: Collision diagrams}. They are analogous to the direct (Hartree) and exchange (Fock) collisions discussed in the BEC theory \cite{proukakis2008finite, morgan2000gapless}. 
We mention that, other than in our case, in open systems, beyond the weakly interacting regime the terms in the interaction operator are accompanied by also those of the energy non-conserving processes, see e.g. \cite{proukakis2008finite, gardiner2000quantum, griffin2009bose}. 

Performing the integrals (\ref{eqn:matrix_element}) for the double-box trap geometry and the corresponding single-particle states $\psi_{0/1}$ from Sec.~\ref{Sec: A single particle in a double-well}, we obtain the matrix elements
\begin{subequations}
\label{eqn:calculated_matrix_elements}
\begin{eqnarray}
    {V}_{0000} &=&  \left(2 - \left[\frac{\Delta E}{\Delta E^a}\right]^2\right){V} +\mathcal{O}(\Delta E/E) \label{eq: V0000a}\\
        {V}_{1111} &=&  \left(2 - \left[\frac{\Delta E}{\Delta E^a}\right]^2\right){V} \label{eq: V1111a}+\mathcal{O}(\Delta E/E)\\
    {V}_{1010} &=& \left[\frac{\Delta E}{\Delta E^a}\right]^2 {V}+\mathcal{O}(\Delta E/E) \label{eq: V1100a}\\
    &=&{V}_{0101}={V}_{1001}={V}_{0110}\nonumber\, ,
\end{eqnarray}
\end{subequations}
which are ingredient to Eq.~(\ref{eqn:interaction_operator}).
Here we neglect all terms of order $\Delta E/E$, i.e., the overlap of the left and right box wave functions.
While this overlap is crucial for the Josephson dynamics, later discussed in this paper, it merely contributes quantitatively to the interparticle interaction, as we show in Appendix~\ref{Appendix: Collisional matrix elements}.

In the Eqs.~(\ref{eqn:calculated_matrix_elements}) 
 we see that the matrix elements are characterized by the ratio $0\leq \Delta E/\Delta E^a\leq 1$ of the energy gaps with and without acceleration $\vec{a}$ and a single parameter
\begin{equation}
    V = \frac{g}{2}\left( \frac{Lm E + \hbar\sqrt{2mE /3}(2 + 1/3 \times E/U_0) }{ (L\sqrt{2mE/3}  + 2\hbar)^2 }\right)^3
    \label{eq:parameter_V}
\end{equation}
determined by the trap geometry, the interaction strength, as well as the single particle's mass and energy.
Note that the quantities entering Eq.~(\ref{eq:parameter_V}) are not independent. 
That is, the energy $E=E(L,U_0,m)$ is fixed by the other quantities and can be obtained numerically, as we do in Appendix~\ref{Appendix: Potential geometry and single-particle states}.

\section{Density matrix and single-particle observables}
\label{sec: Density matrix and single-particle observables}

\subsection{Density matrix evolution}

In this Section we discuss the evolution of the density matrix (\ref{eqn:density_matrix}), following Ref.~\cite{blum2012density}. 
Considering the density matrix in the interaction picture
\begin{equation}
\hat{\rho}_{I}(t) = e^{i \hat{\mathcal{H}}_0 t/\hbar} \hat{\rho}(t) e^{-i \hat{\mathcal{H}}_0 t/\hbar},
\end{equation}
its evolution 
\begin{equation}
    \dot{\hat{\rho}}_I(t) = - \frac{i}{\hbar} [\hat{\mathcal{V}}, \hat{\rho}_I(t)] \, 
    \label{eq: Liouville equation}
\end{equation}
is determined by the interaction operator $\hat{\mathcal{V}} = e^{i \hat{\mathcal{H}}_0 t/\hbar} \hat{\mathcal{V}}e^{-i \hat{\mathcal{H}}_0 t/\hbar}$. The latter preserves its form in the interaction picture, since it is energy-conserving as discussed in Sec.~\ref{Subsec: Collisional interaction}.
Describing the density matrix in the many-particle basis (\ref{eqn:many-particle_eigen_states}),
its matrix elements are 
\begin{eqnarray}
\rho^I_{N_1, N_1'}(t) &=&\rho^0_{N_1, N_1'}e^{-it/\hbar \mathcal{V}_{N_1, N_1'}} \, ,
\end{eqnarray}
where $\rho^0_{N_1, N_1'}=\rho_{N_1, N_1'}(t=0)$ is the initial state of the system, and 
\begin{eqnarray}
    \mathcal{V}_{N_1, N_1'} &=& \frac{V_{0000}}{2}[N_1(N_1 - 2N + 1) - N_1' (N_1' - 2N + 1)] \nonumber \\
    &+& \frac{V_{1111}}{2}[N_1 (N_1 - 1) - N_1' (N_1' - 1)] \label{eqn:another_V} \\ &+& (V_{0101} + V_{1010})[(N - N_1) N_1 - (N - N_1') N_1'] \nonumber
\end{eqnarray}
governs its evolution. 
Transforming this solution of 
Eq.~(\ref{eq: Liouville equation}) back to the Schrödinger picture, we get 
\begin{equation}
\rho_{N_1, N_1'}(t) = \rho^0_{N_1, N_1'} e^{-i(\mathcal{E}_{N_1} - \mathcal{E}_{N_1'} + \mathcal{V}_{N_1, N_1'})t/\hbar} \label{eq: density matrix elements evolution}. 
\end{equation}
Thus, we see that the collisional interaction shifts the energy levels of the system by $\mathcal{V}_{N_1, N_1'}$. Note, due to $\mathcal{V}_{N_1, N_1} = 0$, the diagonal density matrix elements remain constant in time $\rho_{N_1, N_1}(t) = \rho_{N_1, N_1}(0)$, as is evident from the energy and particle number conservation in a closed system.
The off-diagonal elements of the density matrix evolve with 
\begin{eqnarray}
    \hbar \omega_{N_1, N_1'} &=&\mathcal{E}_{N_1'} - \mathcal{E}_{N_1} - \mathcal{V}_{N_1, N_1'}\label{eq: omega ab}\\
    &=& (N_1'-N_1)[\Delta E^a 
    + V^a(N - N_1 - N_1')] \,,\nonumber
\end{eqnarray}
where the interaction now enters via the single parameter
\begin{equation}
        V^a = \left(3 \left[\frac{\Delta E}{\Delta E^a}\right]^2 - 2\right) V \, ,
\end{equation}
emerging from using the explicit form of the matrix elements (\ref{eqn:calculated_matrix_elements}) in the expression (\ref{eqn:another_V}). 
Note that we fulfill $\omega_{N_1', N_1} = -\omega_{N_1, N_1'}$, as immediately stems from the fact that the density matrix is Hermitian.

\subsection{Effective density matrix}
\label{Subec: Effective density matrix}

In the last section, we derived the density matrix $\hat{\rho}(t)$ that contains complete information about the many-particle system and its evolution. 
Our ultimate goal, however, is to describe Josephson dynamics 
and, in particular the observable oscillations of particle populations between the wells of a double-well potential.

These populations are expectation values of the number operators
\begin{equation}
\hat N_{L/R} = \underbrace{\hat{P}_{L/R} \oplus \mathbb{1} \oplus \dots \oplus \mathbb{1}}_{N\textnormal{-times}} + \dots + \mathbb{1} \oplus \mathbb{1} \oplus \dots \oplus \hat{P}_{L/R} \, , \label{eqn:NLR}
\end{equation}
which are constructed from the projectors $\hat{P}_{L/R}=|\psi_{L/R}\rangle\langle\psi_{L/R}|$, projecting on the state of a single particle in the left or right well, respectively. 

As we discuss in detail in Ref.~\cite{PhysRevA.109.043321}, the expectation value 
\begin{equation*}
    \langle \hat N_{L/R}\rangle_{\rho}  = \mathrm{tr}(\hat{\rho}\hat N_{L/R}) = \mathrm{tr}(\hat{\rho}_\mathrm{e}\hat{P}_{L/R})
\end{equation*}
can be evaluated when only the effective density matrix $\hat \rho_\mathrm{e}$ is known.
In our case, having that only two single particle energy eigenstates $\psi_{0/1}$ are considered, $\hat \rho_e$ is a $2\times2$ matrix.
It coincides with the reduced density matrix of a single particle in a bath of all the other $N - 1$ bosons \cite{erdahl2012density} and can therefore be interpreted as a description of the many-particle system by an average boson. 

Using the solution (\ref{eq: density matrix elements evolution}) for $\hat{\rho}(t)$,  we obtain the \changesOK{effective density matrix
\begin{equation}
\hat{\rho}_\mathrm{e} = \sum_{ij} \alpha_{ij} |\psi_i \rangle \langle \psi_j | \label{eqn:eff_dens_matrix}
\end{equation}
in the $\psi_{0/1}$  basis with the elements
}
\begin{subequations}
\label{eq: alpha0011}
    \begin{eqnarray}
    \alpha_{00} &=& \sum_{N_1 = 0}^N \rho_{N_1, N_1}^0 (N - N_1)\\
    \alpha_{11} &=& \sum_{N_1 =0}^N \rho_{N_1, N_1}^0 N_1 \\
    \alpha_{10}(t) &=& \sum_{N_1 = 0}^{N-1} \sqrt{(N - N_1)(N_1 + 1)} \nonumber \\
    & &\, \times \rho_{N_1, N_1 + 1}^0 e^{i \omega (N_1)t} \\
    \alpha_{01}(t) &=&\alpha_{01}^
    \star(t) ,
\end{eqnarray}
\end{subequations}
which are determined by the diagonal elements $\rho^0_{N_1, N_1}$ and the nearest off-diagonal elements 
 $\rho^0_{N_1, N_1 + 1}$ of the initial density matrix. \changesOK{Note, that in the case of a fully coherent configuration $|\alpha_{10}| = \sqrt{\alpha_{00}\alpha_{11}}$, while in the general case we have $|\alpha_{10}| \in [0, \sqrt{\alpha_{00}\alpha_{11}}$] \cite{PhysRevA.109.043321}.} Moreover, the evolution is governed by the frequency
\begin{equation}
    \omega(N_1) : = \omega_{N_1, N_1 + 1} = \frac{1}{\hbar}\left[\Delta E^a + V^a (N - 2N_1 - 1)\right]
\end{equation}
of transitions between the many-particle states $|N_1\rangle$ and $|N_1\pm1\rangle$, cf. Eq.~(\ref{eq: omega ab}).
Other transitions  $|N_1\rangle\to |N_1\pm 2\rangle$, or higher, do not appear, since we only consider projectors $\hat{P}_{L/R}$ acting on a state of each single boson independently, cf. Eq.~(\ref{eqn:NLR}).

To discuss Josephson dynamics,
we need to transform the effective density matrix 
$\hat{\rho}_{e LR} = \hat{T}(\xi)\hat{\rho}_{e}\hat{T}(\xi)^{-1}$ 
from the $\psi_{0/1}$-basis to the basis of left and right states, using the transformation matrix $\hat{T}(\xi)$  from Eq.~(\ref{eq:left-right states}).
After this transformation,  we can identify the diagonal elements of
\begin{equation}
    \hat{\rho}_{e LR}(t) =  \begin{bmatrix} 
	N_L (t) & 	A(t)  \\
        A^*(t) & 	N_R (t)  \\ 
	\end{bmatrix} \label{eq: effective density matrix}
\end{equation}
with the time-dependent occupation numbers
\begin{subequations}
\label{eq: NLR population imbalances}
    \begin{eqnarray}
    N_{L}(t) &=& \alpha_{00}\cos^2 \xi + \alpha_{11} \sin^2 \xi \\
    & & \hspace{6em} +  \textrm{Re}[\alpha_{10}(t)] \sin(2\xi) \nonumber\\[0.5em]
     N_{R}(t) &=& \alpha_{00}\sin^2 \xi + \alpha_{11} \cos^2 \xi \\
    & & \hspace{6em} 
     -  \textrm{Re}[\alpha_{10}(t)] \sin(2\xi) \nonumber
\end{eqnarray}
\end{subequations}
of the wells.
The also introduced off-diagonal element
\begin{eqnarray}
    A(t) &=& \frac{\alpha_{00} - \alpha_{11}}{2} \sin(2 \xi) \label{eq: A paramter}\\
    & &\qquad - \textrm{Re}[\alpha_{10}(t)] \cos (2\xi) + i \textrm{Im}[\alpha_{10}(t)] \nonumber \, 
\end{eqnarray}
and its conjugated will allow us to quantify the coherence between the two spatially separated subsystems, constituting the left and the right well states. These
two subsystems now can be seen as two coupled BECs -- a picture, that is often stressed in literature \cite{Smerzi1997}.

\subsection{
Population imbalance and degree of coherence}
\label{subsec: Population imbalance and degree of fragmentation}

Having found the explicit form (\ref{eq: NLR population imbalances}) for the population numbers $N_{L/R}(t)$ of each well, we now can construct the observable population imbalance 
\begin{eqnarray}
    Z(t) &=& (N_L (t) - N_R (t))/N \nonumber\\
    &=&
    \frac{\cos (2\xi)}{N}\sum_{N_1 = 0}^N \rho^0_{N_1, N_1} (N - 2N_1)  \label{eq: gen sol for pop imb}\\
    && + \,\frac{\sin (2\xi)}{N} \sum_{N_1 = 0}^{N-1}\sqrt{(N_1 + 1)(N - N_1)} \nonumber \\
    & & \quad\times\!\left( \rho_{N_1, N_1 + 1}^0 e^{i \omega (N_1) t} + \rho_{N_1+1, N_1}^0 e^{-i \omega (N_1) t}  \right) \, \nonumber,
\end{eqnarray}
which for our closed system is the only degree of freedom since $N_L(t) + N_R(t)=N$ gives the total particle number.
The first term in Eq.~(\ref{eq: gen sol for pop imb}) 
is the average population imbalance, which gives a contribution in the presence of acceleration $\mathbf{a}\neq 0$
and vanishes for $\mathbf{a}= 0$.
The second, time-dependent, term 
gives rise to the Josephson oscillations. 

For a specific scenario, i.e., a given initial condition $\rho^0_{N_1, N_1'}$, the imbalance $Z(t)$ encodes the entire observable dynamics of the system.
However, further quantities can be introduced to characterize the behavior of $Z(t)$.
In particular, we are interested in a parameter to quantify the \emph{decoherence} of the system due to interparticle collisions, causing the oscillations to die off after a certain time.
To obtain this parameter,
we bring the density matrix (\ref{eq: effective density matrix}) into a diagonal form
\begin{equation}
S\hat{\rho}_{e LR}(t)S^{-1}
= \frac{N}{2}\begin{bmatrix} 
	1+f(t) & 	0  \\
        0 & 	1-f(t)  \\ 
	\end{bmatrix} \label{eqn:diag_DM}\,,
\end{equation}
where $f(t)$ ranges from $0$ to $1$ and is given by
\begin{equation}
    f(t) = \sqrt{Z^2(t) + \left(\frac{2|A(t)|}{N}\right)^2} \label{eq: f parameter}.
\end{equation}
While here, we don't want to specify the transformation $S(t)$ or the resulting time-dependent basis states $|s_1(t)\rangle$ and $|s_2(t)\rangle$ implied by Eq.~(\ref{eqn:diag_DM}), we see that for $f=1$ we obtain $S\hat{\rho}_{e LR}(t)S^{-1}=\mathrm{diag}(N,0)=N|s_1\rangle\langle s_1|$. This means that the system is in a single state $|s_1(t)\rangle = c_1(t) |\psi_L\rangle + c_2(t) |\psi_R\rangle$ that can be written as a superposition of the $|\psi_{L/R}\rangle$ states. Thus, for $f=1$ the system is described by a fully coherent state. 
In the opposite case, $f=0$, the system is in an incoherent mixture 
$\frac{N}{2}|s_1\rangle\langle s_1|+\frac{N}{2}|s_2\rangle\langle s_2|$.

Based on this discussion, it is apparent that the parameter $f\in[0,1]$ has the meaning of a \emph{degree of coherence}. In our previous work \cite{PhysRevA.109.043321}, we discussed this parameter as constant in time for non-interacting systems. 
In what follows, the analysis of $f(t)$ together with $Z(t)$ will allow us to investigate the process of decoherence during Josephson oscillations in an interacting Bose gas.

\section{Josephson effect}
\label{Sec: Decoherence}

In the last Section we have obtained the population imbalance $Z(t)$ and the degree of coherence $f(t)$, that 
can be utilized to describe Josephson oscillations of a weakly self-interacting BEC in a double-well potential. 
We now take this general result and apply it to the specific case, where all $N$ particles initially occupy just a single (e.g., the left) well of the potential.
This describes a standard experimental scenario, where the BEC is initially loaded into a single well, using modern trap manipulation techniques \cite{Albiez_2005}.
In our model, this gives rise to the 
initial state of the system
\begin{eqnarray}
    | \mathcal{L} \rangle &=& \frac{1}{\sqrt{ N!}}(\cos \xi \hat{a}_0^\dagger + \sin \xi \hat{a}_1^\dagger)^N|0\rangle \, ,\label{eq: initial left well state}
\end{eqnarray}
where the operators $\hat{a}_{0/1}^\dagger$ create particles in the ground or excited state, respectively. Thus, according to Eq.~(\ref{eq:left-right states}) the superposition $\cos \xi \hat{a}_0^\dagger + \sin \xi \hat{a}_1^\dagger$ creates a boson in the left-well state $\psi_{L}$. 
The corresponding initial condition for the density matrix reads
\begin{eqnarray}
    \hat{\rho}^0 &=& |\mathcal{L} \rangle \langle \mathcal{L} | \label{eq: left state}  \\
    = &N!& \sum_{N_1=0}^{N} \sum_{N_1' =0}^N \frac{(\cos \xi)^{2N - N_1 - N_1'} (\sin \xi)^{N_1+N_1'}}{\sqrt{N_1! N_1'! (N - N_1)! (N - N_1')!}} |N_1\rangle \langle N_1'|\,, \nonumber
\end{eqnarray}
where we sum over all many-particle energy eigenstates $|N_1\rangle$ and $|N_1'\rangle$, which are defined in Eq.~(\ref{eqn:many-particle_eigen_states}).

\subsection{Josephson dynamics without external acceleration}
\label{subsec: Josephson dynamics without external acceleration}
In what follows, we want to focus on the effect of decoherence due to collisional interactions. 
Therefore, here we first consider Josephson oscillations in the absence of external acceleration.

To find the fractional population imbalance and the degree of coherence, we revisit Eqs.~(\ref{eq: gen sol for pop imb}) and (\ref{eq: f parameter}) using the $\rho^0_{N_1,N_1'}$ that can be read of Eq.~(\ref{eq: left state}). In both equations the case of $\vec{a}=0$ is parameterized by $\xi=\pi/4$.
This yields
\begin{eqnarray}
     Z(t) &=& \left(\cos \left[\frac{V}{\hbar}t \right] \right)^{N-1}\cos \left[ \frac{\Delta E}{\hbar}  t \right]\, ,\label{eq: left well popul imbalance}\\
     f(t) &=& \bigg|\cos \left[\frac{V}{\hbar}t\right]\bigg|^{N - 1} \label{eqf(t)}.
\end{eqnarray}
Starting at $Z(0)=1$ ($N_L(0)=N$), Eq.~(\ref{eq: left well popul imbalance}) shows harmonic oscillations with frequency $\Delta E/\hbar$, as would be expected in the non-interacting case. 
In the presence of collisional interactions, which enter our 
formalism via the parameter $V$ from Eq.~(\ref{eq:parameter_V}),
these oscillations are modified by an additional, periodic function $\cos^{N-1}(Vt/\hbar)$. The absolute value of this function coincides with the degree of coherence (\ref{eqf(t)}).

\begin{figure}[t]
\centering
\includegraphics[width=.45\textwidth]{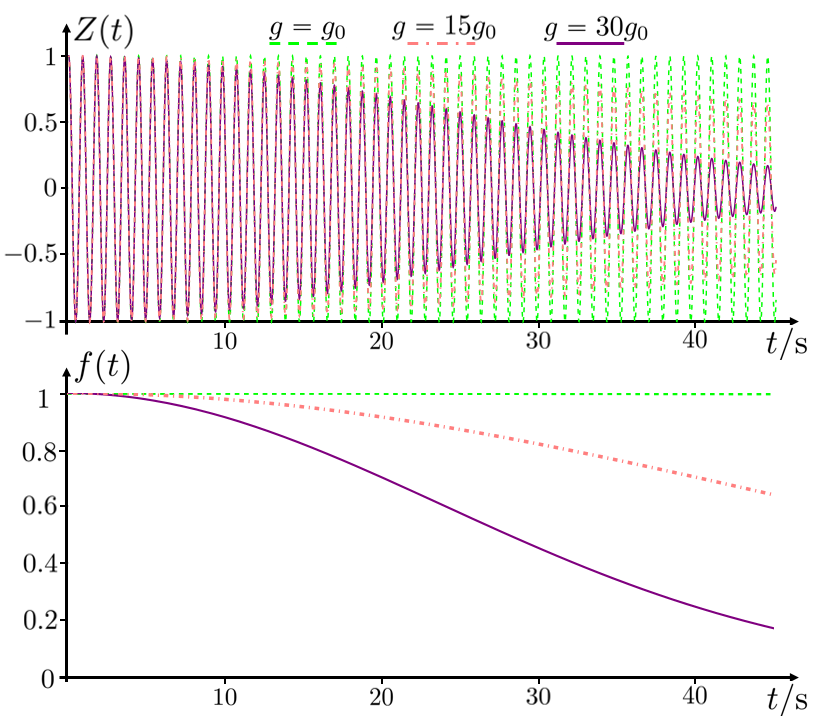}
   \caption{Population imbalance $Z$ (top) and degree of coherence $f$ (bottom) as  functions of time $t$. Here the number of bosons is $N = 10^4$ and $g_0 = 6.86 \times 10^{-55}$~J$\times$m$^3$.}
   \label{Fig: Collisional decoherence}
\end{figure}

In the considered regime of weak interactions ($V\ll \Delta E$) the period $\sim\hbar/V$ is much longer than the time scale $\hbar/\Delta E$ of Josephson dynamics in Eq.~(\ref{eq: left well popul imbalance}). 
Thus, for times $t \lesssim \hbar/V$ the degree of coherence appears as 
a damping factor for the Josephson oscillations.
The decrease of $f(t)$ with time is a signature of the decoherence of the initially coherent state (\ref{eq: left state}), c.f. Ref.~\cite{pitaevskii2016bose}.
Moreover, we find that after some time $t 
\gtrsim \hbar/V$ 
the amplitude of Josephson oscillations in Eq.~(\ref{eq: left well popul imbalance})  is revived to its initial value, implying that the coherence is periodically restored.
This stems from the fact that in a closed system with conserved energy and entropy the initial system state is never forgotten.
On a technical level, this can be understood from the density matrix elements $\rho_{N_1, N_1'}(t)$, each evolving with its own phase, as displayed in Eq.~(\ref{eq: density matrix elements evolution}).
These matrix elements interfere destructively, until some time, given by the largest period $\sim \hbar/V$ involved, is reached. 
Hence, the revival of the coherence 
is a consequence of the discrete, finite energy spectrum (\ref{eq: omega ab}), see also Refs.~\cite{pitaevskii2016bose,castin1997relative, PhysRevLett.77.2158}.

In experiment, however, the assumption that a BEC is a closed system typically does not hold over time scales $\sim \hbar/V$, due to the significant exchange of particles and energy with 
the environment. 
This provides an additional channel for decoherence, leads to a finite BEC lifetime \cite{griffin2009bose}, and prevents a revival of the oscillations.

To illustrate the Josephson dynamics and its decoherence, 
described by Eqs.~(\ref{eq: left well popul imbalance}) and (\ref{eqf(t)}) for a realistic experimental scenario, we consider a BEC of $^{39}$K atoms with a small scattering wavelength in the range $a_s = 0.006\, a_0 -  0.18\, a_0$, as realized in Refs.~\cite{masi2021multimode,fattori2008atom}. Here $a_0$ is the Bohr radius.
The interaction constant $g = 4\pi \hbar^2 a_s/m$ of such a gas varies from $g_0 = 6.86 \times 10^{-55}$ $\textrm{kg} \times \textrm{m}^5/\textrm{s}^2$ to $30g_0 = 2.06 \times 10^{-53}$ $\textrm{kg} \times \textrm{m}^5/\textrm{s}^2$. 
Moreover, we assume a typical BEC of $N = 10^4$ bosons in our trap geometry (revisit Fig.~\ref{Fig: trap geometry}) of size $L = 3.68 \mu$m (concentration $10^{14}$ $\textrm{cm}^{-3}$) and a depth $U_0=3.16 \times 10^{-32}$~J of the potential. For such a setup we obtain Josephson oscillations in a time scale of $\hbar/\Delta E = 0.14$~s, while the collisional decoherence time scale $\hbar/V\sim g^{-1}$ ranges from $7.66 \times 10^4$~s to $2.55 \times 10^3$~s.

In the Fig.~\ref{Fig: Collisional decoherence} we show the decay of Josephson oscillations (\ref{eq: left well popul imbalance}) for different interaction strengths $g$.
For larger $g$ (see $g=15g_0$ or $30g_0$ in the Figure) we observe a significant decay of the oscillation amplitude, in agreement with the findings of Ref.~\cite{fattori2008atom}, where the collisional interaction was observed to be the main source of decoherence, and Ref.~\cite{sakmann2014universality}, where similar decoherence patterns were obtained numerically from Bose-Hubbard model.
In contrast, for $g = g_0$, as given in Ref.~\cite{masi2021multimode}, the effect of collisional interactions is very small 
and leads to a $10^{-5}$ decrease of the amplitude on the time scale of $10$ oscillation periods. Thus, this case is almost indistinguishable from the Josephson dynamics of a non-interacting BEC.

\subsection{Impact of acceleration}
\label{SubSec: Impact of acceleration}
Having discussed the collisional decoherence in the absence of acceleration, we now analyze the interplay of both effects.
Considering again the scenario (\ref{eq: left state}) where all 
bosons initially
occupy the left well, 
the population imbalance (\ref{eq: gen sol for pop imb}) in the presence of external acceleration $\mathbf{a}$ reads
\begin{eqnarray}
   && Z(t) =  \cos^2 (2\xi)
    + 4 \cos^{2N} \xi \sin^2 \xi \label{eqn:Zoftwithchi}\\ &&\times \textrm{Re} \left[e^{i(\Delta E^a + V^a(N-1))t/\hbar}(1 + e^{-2i V^a t/\hbar}\tan^2 \xi)^{N-1}\right]\, \nonumber.
\end{eqnarray}
Note that for $\xi=\pi/4$ the population imbalance (\ref{eq: left well popul imbalance}) in the absence of acceleration is reproduced. 
Eq.~(\ref{eqn:Zoftwithchi}) covers the more general case, where $\xi$ for a given acceleration can be determined from the double-well geometry as described in Sec. \ref{Sec: A single particle in a double-well}.
\begin{figure}[htp]
    \centering  \includegraphics[scale = 0.45]{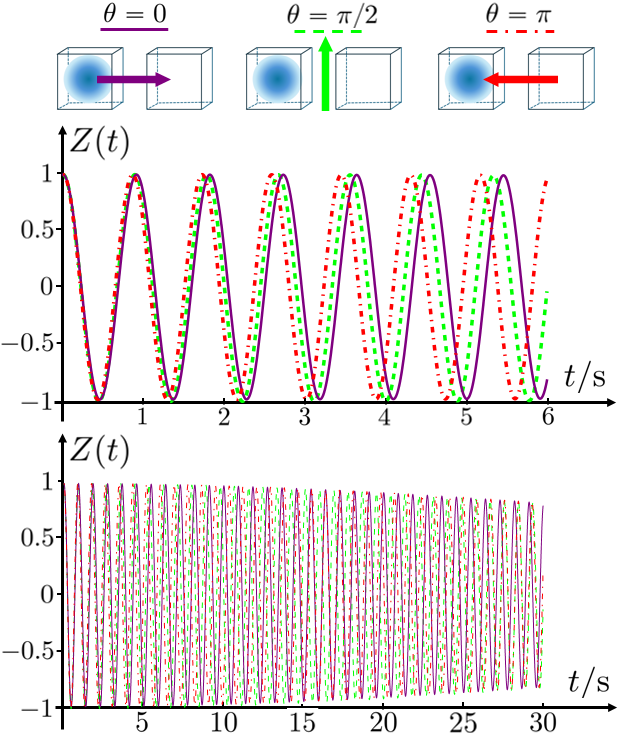}
    \caption{Top: the pairs of cubes depict the trap geometry (\ref{Fig: trap geometry}), while the blue cloud illustrates the initial condition (\ref{eq: left state}). The arrows show the three cases of the direction of the applied acceleration $\mathbf{a}$. Middle and bottom: Population imbalance $Z$ as a function of $t$ for different time scales. Here the number of bosons is $N = 10^4$ and self-interaction strength  $g = 15 g_0 = 1.03 \times 10^{-53}$~J$\times$m$^3$ is the same for all graphs. The purple (solid), green (dashed) and red (dot-dashed) lines correspond to the cases of acceleration $\mathbf{a}$ directed along $\theta =0 $, $\pi/2$, $\pi$ respectively, cf. upper figure. }
    \label{Fig: Collisional decoherence with acceleration}
\end{figure}
As an example, we consider
an acceleration of $a = 10^{-5}a_\Earth$ (where $a_\Earth = 9.81\, \textrm{m}/\textrm{s}^2$)
applied to a BEC of $^{39}$K atoms as considered in Sec.~\ref{subsec: Josephson dynamics without external acceleration} ($L = 3.68 \mu\mathrm{m}$, $U_0= 3.16 \times 10^{-32}$~J).
For the acceleration aligned with ($\theta=0,\pi$) or
perpendicular to ($\theta=\pi/2$) the trap we obtain
\begin{eqnarray*}
\xi(\theta=0)\phantom{/2}  &= & \pi/4+\pi/68 \, ,\\
    \xi(\theta=\pi/2)  &= & \pi/4 \, ,\\ \xi(\theta=\pi)\phantom{/2}  &= & \pi/4-\pi/68\,.
\end{eqnarray*}
Evaluating Eq.~(\ref{eqn:Zoftwithchi}) for those particular values gives rise to Josephson dynamics, as shown in Fig.~\ref{Fig: Collisional decoherence with acceleration}. 
We see that the result for $\theta=\pi/2$ parametrically coincides with the case of $a=0$, such that perpendicular acceleration has no effect on the Josephson junction.

As seen from Fig.~\ref{Fig: Collisional decoherence with acceleration}, the dominating effect of acceleration is the change of the oscillation period, while the decay of the oscillation amplitude is nearly unaffected. 
As discussed in Sec.~\ref{subsec: Josephson dynamics without external acceleration}, the latter is determined by the degree of coherence, which in the presence of acceleration is given by \begin{eqnarray}
    f(t) = \sqrt{\cos^2(2\xi) + \sin^2(2 \xi)  (R(t)\cos^4 \xi)^{N-1}} \label{eq: general f(t)}\\
    R(t) = 1 + 2 \cos \left(\frac{2V^a}{\hbar}t \right)\tan^2 \xi + \tan^4 \xi \nonumber \,. 
\end{eqnarray}
Expanding this expression for small values of $a$, we find 
\begin{eqnarray}
    f(t) = \bigg| \cos \left[\frac{V}{\hbar}t \right] \bigg|^{N-1}+   \mathcal{O}\left[\left(\frac{mLa}{\Delta E}\right)^2\right]\,,
    \label{eq: deg of coh small acc}
\end{eqnarray} 
which indeed, up to linear order, coincides with the result (\ref{eqf(t)}) obtained for $a=0$.

In what follows we investigate how the change of Josephson dynamics due to acceleration can be used as a basis for a BEC double-well accelerometer \cite{roy2025assessing, sacchetti2016accelerated, tonel2005behaviour}.

\subsection{Sensitivity of a BEC double-well accelerometer}
We want to investigate the feasibility of the setup to sense small accelerations $a \ll \Delta E/(Lm)$ by evaluating the period of the first few Josephson oscillations ($t \ll \hbar/V$). In this regime, Eq.~(\ref{eqn:Zoftwithchi}) simplifies to
\begin{eqnarray}
    Z(t) &\approx& \left(\cos \left[\frac{V}{\hbar}t \right] \right)^{N-1} \label{eq:Zfor smallacc} \\ &\times& \cos \left[\left(\Delta E - 2(N-1)\frac{V}{\Delta E} \frac{\chi (\theta)}{L} L m a \right)\frac{t}{\hbar}\right] \nonumber \, ,
\end{eqnarray}
where $\chi(\theta) \approx 3L \cos \theta/2$.
Based on Eq.~(\ref{eq:Zfor smallacc}) we can make a general conclusion about the sensitivity of the system to the external acceleration. 
For that purpose, we derive the change of oscillation period $\delta T=\frac{\partial T}{\partial a} \delta a$ with acceleration, giving
\begin{equation}
    \delta T = 6\pi \hbar \frac{ N V  L m}{\Delta E^3} |\cos \theta| \delta a \, ,
    \label{eq: delta T presicion}
\end{equation}
where we accounted for $N - 1 \approx N$.
The currently achievable precision of period measurement in experiments \cite{Albiez_2005,PhysRevLett.118.230403} is $\delta T/T  =\frac{\Delta E}{2\pi\hbar}\,\delta T \sim 10^{-2}$. Thus, for the example setup introduced in Sec.~\ref{subsec: Josephson dynamics without external acceleration}, and  $g=g_0$, we obtain a potential resolution of 
$\delta a \sim 5 \times 10^{-4} a_\Earth$ for  acceleration measurements, which was achieved in experiments with non-interacting BECs, e.g. in Ref.~\cite{masi2021multimode}. Eq.~(\ref{eq: delta T presicion}) tells us, how this sensitivity can be improved, for instance, by increasing the depth $U_0$ of the trap, which would
rapidly decrease $\Delta E$ and increase $V$, as shown in the Appendices~\ref{Appendix: Potential geometry and single-particle states} and \ref{Appendix: Collisional matrix elements}, respectively.
With that, already considering $U_0\to 2U_0$, provides a potential way to reach $\delta a \sim 10^{-6} a_\Earth$ while keeping the relative precision $\delta T/T = 10^{-2}$ of period measurement fixed. 
Sensitivities of $\delta a \sim 10^{-6} a_\Earth$ and beyond are planned to be achieved with the experimental setups using BEC interferometry, as suggested in Refs.~\cite{gersemann2020differential, PhysRevLett.117.203003}.

The discussed acceleration measurement scheme relies on the linearity of the relation (\ref{eq: delta T presicion}), which only holds up to a certain  energy scale of 
interactions $NV$.
Beyond that, the
Josephson oscillations are expected to become anharmonic, or even reach other dynamical regimes in which the relation between $\delta T$ and $\delta a$ may tremendously differ from Eq.~(\ref{eq: delta T presicion}).
\changesOK{For instance,
macroscopic quantum self-trapping becomes possible when the interactions are strong enough \cite{tonel2005behaviour, links2006two}.}
In what follows, we investigate
the limits, which the energy scale $NV$ sets on the applicability of our model.

\section{Comparison with the standard two state model}
\label{Sec: Comparison with the standard two state model}

For the considered scenario of a BEC with weak interparticle interactions, the decoherence is a slow process compared to the time scale of Josephson dynamics, see Figs.~\ref{Fig: Collisional decoherence}-\ref{Fig: Collisional decoherence with acceleration}.
Therefore, within the timescale of the first few Josephson oscillations, the degree of coherence
can be approximately set to $f(t) = 1$.
While this approximation does not account for the decoherence effects, we can test our prediction (\ref{eq: delta T presicion}) for the shift of the oscillation period in a simpler framework, which accounts for the interactions non-perturbatively.
The assumption $f(t) = 1$ leads to the pure state model, suggested in Refs.~\cite{Smerzi1997, raghavan1999coherent}, where the whole many-particle system is described by a single state wave function $\Psi$, which obeys a Gross-Pitaevskii equation
\begin{equation}
    i\hbar \frac{\partial \Psi}{\partial t} = \left[-\frac{\hbar^2}{2m}\nabla^2 + U(\mathbf{r}) + \delta U(\mathbf{r}) + g|\Psi|^2\right]\Psi \label{eq: GP}\,.
\end{equation}
In our case, $U(\mathbf{r})$ stands for the double-well trap geometry and $\delta U(\mathbf{r}) = maz\cos\theta + max \sin\theta$ for the additional potential due to acceleration $\mathbf{a}$, as introduced in  Sec.~\ref{Sec: A single particle in a double-well}. Following Refs.~\cite{Smerzi1997, raghavan1999coherent}, the pure state for the double-well geometry can be approximated by 
\begin{equation}
\Psi = \sqrt{N_L(t)}e^{i\varphi_L}\psi_L + \sqrt{N_R(t)}e^{i\varphi_R}\psi_R\,, \label{eqn:order_parameter}
\end{equation}
where $\psi_{L/R}$ are the left/right well states from Eq.~(\ref{eqn:psi_LR}).
Plugging this ansatz in Eq.~(\ref{eq: GP}) and integrating the result with the left and right state $\int d\mathbf{r} \psi_{L/R} \times$ leads to a system of two coupled differential equations 
\begin{subequations}
\label{eq: main Jos eqs}
\begin{eqnarray}
     \dot{Z} &=& \frac{\Delta E}{\hbar} \sqrt{1 - Z^2}\sin \changesOK{\varphi} \\
   \changesOK{\dot{\varphi}} &=& -\frac{\Delta E}{\hbar} \left(\Xi + \Lambda Z + \frac{Z}{\sqrt{1 - Z^2}}\cos \phi\right) \, ,
\end{eqnarray}
\end{subequations}
which were introduced as \emph{Josephson equations}
for the population imbalance $Z(t) = (N_L(t) - N_R(t))/N$ and the phase difference $\varphi(t) = \varphi_R(t) - \varphi_{L}(t)$ in Refs.~\cite{Smerzi1997, raghavan1999coherent}.
In our case,
the parameters $\Xi = -3aLm \cos \theta/\Delta E$ and $\Lambda = -2VN/\Delta E$ characterize the impact of the acceleration and the interaction strength, respectively.
For more details on the derivation, see Appendix~\ref{Appendix: Comparison with the pure state approach}. 
In the regime of weak interactions ($|\Lambda| < 1$) the solution of Eqs.~(\ref{eq: main Jos eqs}) can be well approximated by harmonic oscillations with the frequency $\Delta E(1 - \Lambda \Xi/2)/\hbar$, as we show in Appendix~\ref{Appendix: Comparison with the pure state approach}. The shift $\Delta E\Lambda \Xi/(2\hbar) = 3 N V L m a \cos \theta/(\hbar\Delta E)$ is the same as in Eq.~(\ref{eq:Zfor smallacc}).
For larger interaction strengths $|\Lambda|\gtrsim 1$ 
nonlinear effects become significant and the solution of (\ref{eq: main Jos eqs}) exhibits non-harmonic dynamics as discussed in Ref.~\cite{raghavan1999coherent}.
Hence, \changesOK{$|\Lambda|\lesssim 1$}, i.e.,  $2VN < \Delta E$, sets an important constraint for the applicability of our model.
For the example setup considered in this paper, this corresponds to  $g < 30g_0$.

\section{Comparison with the phase fluctuations approach}
\label{Sec:Density matrix and phase fluctuations approach}

The density matrix approach, discussed in the previous sections, is not the only way to theoretically predict and describe decoherence in a Josephson junction. For instance, in Refs.~\cite{PhysRevA.55.4318, PhysRevLett.87.180402} a quantized version of the Josephson equations \changesOK{(\ref{eq: main Jos eqs})}, where the variables $Z(t)$ and $\varphi(t)$ were replaced by the corresponding operators, was
used.
\changesOK{In that formalism,  the phase of the entire condensate can be measured sharp,
if the gas is in an eigenstate of the phase operator. 
This means that
the Bose gas is in a fully coherent state \cite{PhysRevA.55.4318}.
Otherwise, if the expectation value of the phase operator with respect to the state 
is not sharp, its 
\emph{fluctuations} are a measure of decoherence \cite{pitaevskii2016bose,castin1997relative, RevModPhys.90.035005}.}

 In the following we will show how these phase fluctuations are connected to the density matrix (\ref{eqn:density_matrix}) of a many-particle bosonic system. Following Ref.~\cite{pitaevskii2016bose}, we start with a fully coherent $N$-particle configuration, described by 
\begin{eqnarray}
|\Phi \rangle &=& \frac{1}{\sqrt{N!}} \left(\hat a^\dagger\right)^N |0 \rangle \, ,\label{eq: Phi state}
\end{eqnarray}
where each boson is in the same single-particle state $|\psi\rangle = \hat{a}^\dagger |0 \rangle $.
\changesOK{In our model, this single-particle state is constructed from the lowest two energy eigenstates $\psi_{0/1}$  of a double-well, see Eq.~(\ref{eq: perturbed states}).}  In the non-interacting case, when the collisions and collisional decoherence are absent, this single-particle wave function is described by
\begin{eqnarray}
    |\psi \rangle &=& \sqrt{\frac{\alpha_{00}}{N}}e^{-iE_0 t/\hbar} |\psi_0 \rangle + \sqrt{\frac{\alpha_{11}}{N}}e^{-iE_1 t/\hbar} |\psi_1 \rangle \nonumber \\
    &\sim& \sqrt{\frac{\alpha_{00}}{N}} |\psi_0 \rangle + \sqrt{\frac{\alpha_{11}}{N}}e^{-i \omega t} |\psi_1 \rangle \label{eq: psi pure}\,,
\end{eqnarray}
\changesOK{where we adopted the notation $\alpha_{00/11}$ from Eq.~(\ref{eq: alpha0011}) to denote the constant populations of ground and exited state, respectively. All time-dependence is governed by the relative phase of the energy eigenstates, which grows linearly with time and is determined by the frequency $\omega=(E_1-E_0)/\hbar$.}
The effective density matrix corresponding to the state $|\psi \rangle$ reads
\begin{equation}
\hat{\rho}^{\mathrm{pure}}_{\mathrm{e}}(\omega) = N|\psi \rangle \langle \psi|= \begin{bmatrix} 
	\alpha_{00} & 	\sqrt{\alpha_{00} \alpha_{11}}e^{i \omega t}  \\
         \sqrt{\alpha_{00} \alpha_{11}} e^{-i \omega t} & 	\alpha_{11}  \\ 
	\end{bmatrix}
    \label{eqn:purestate_rho_fixed_omega}
\end{equation}
and for fixed $\omega$ describes a pure state configuration for all times $t$.

\changesOK{In order to investigate the fluctuations of the phase $\varphi$ between the potential wells, it is sufficient to first consider this problem in the basis of energy eigenstates in which the Eqs.~(\ref{eq: psi pure}-\ref{eqn:purestate_rho_fixed_omega}) are given. For that, we assume that $\omega \in (-\infty, +\infty)$ is not fixed but follows a probability distribution $P(\omega)$ that characterizes a mixture of pure states.
As we will show, the average
\begin{equation}
\hat{\rho}_{\mathrm{e}} = \int_{-\infty}^{+ \infty} d\omega \,\hat{\rho}^{\mathrm{pure}}_{\mathrm{e}}(\omega)\, P(\omega) \label{eqn:average_of_rho_e}
\end{equation}
for a particular $P(\omega)$ coincides with our 
effective density matrix $\hat{\rho}_{\mathrm{e}}$ of the (partially) incoherent system from Eq.~(\ref{eqn:eff_dens_matrix}). 
Considering the off-diagonal element $\alpha_{01}(t)$ of  $\hat{\rho}_{\mathrm{e}}$ from the stated equality (\ref{eqn:average_of_rho_e}), we obtain} 
\begin{equation}
\frac{\alpha_{01}(t)}{\sqrt{\alpha_{00} \alpha_{11}}} = \int_{-\infty}^{+ \infty} d\omega \, e^{i \omega t} P(\omega) \label{eq: alph01phfluct}\, .
\end{equation}
\changesOK{That is, the elements $\alpha_{ij}$ of the effective density matrix coincide with the Fourier transform of 
$P(\omega)$.
Consistently, in the fully coherent case (\ref{eqn:purestate_rho_fixed_omega}), where 
$P(\omega)$ is a delta distribution, the relation $|\alpha_{01}(t)| = \sqrt{\alpha_{00}\alpha_{11}}$ holds. In the general case of (partially) incoherent systems with $|\alpha_{01}(t)| < \sqrt{\alpha_{00}\alpha_{11}}$, Eq.~(\ref{eq: alph01phfluct}) gives the link between the density matrix approach and the phase fluctuations approach.
We, therefore, can use the inverse Fourier transform to determine
\begin{equation}
P(\omega) = \frac{1}{2\pi\sqrt{\alpha_{00} \alpha_{11}}}\int_{-\infty}^{+\infty} dt \,e^{-i \omega t} \,\alpha_{01}(t)\,.
\end{equation}}

\changesOK{Let us illustrate this idea for the weakly interacting BEC accelerometer, discussed in Sec.~\ref{SubSec: Impact of acceleration}. 
Using the elements $\alpha_{ij}$ from Eq.~(\ref{eq: alpha0011}), together with the initial condition (\ref{eq: left state}), we obtain 
\begin{eqnarray}
P(\omega) &=& 
 \hbar \left(\sin \xi\right)^{2(N-1)} \sum_{k=0}^{N-1} C^k_{N-1}\cot^{2k}\xi \label{eqn:P_omega_accelerometer} \\
 & & \qquad\times\, \delta(\hbar \omega - \Delta E^a - V^a (2k-N+1))\,.\nonumber
\end{eqnarray}}
\changesOK{With this distribution, observables for the many-particle state  described by $\hat{\rho}$ 
can be obtained by weighting their pure state expectation values, instead.
In particular, by averaging the population imbalance $Z_{\mathrm{pure}}(\omega t)$, we find
\begin{equation}
 Z(t)  =\frac{1}{N}\mathrm{tr}[\hat{\rho}(\hat N_{L}-\hat N_{R})] = \int d\omega\, P(\omega)\, Z_{\mathrm{pure}}(\omega t)\,, \label{eqn:pop_imbalance}
\end{equation}
which recovers the population imbalance (\ref{eqn:Zoftwithchi})
for the partially incoherent system.
The explicit calculation can be found in Appendix~\ref{Appendix: Density matrix and phase fluctuation},
where we also calculate the
variance of $\omega=(E_1-E_0)/\hbar$, which reads
\begin{eqnarray}
\langle \Delta \omega^2 \rangle &=&
(N-1)\left(\frac{V^a}{\hbar}\right)^2 \sin^2 \left(2\xi\right) \label{eq: variance of omega} \\
&=& (N-1)\left(\frac{V}{\hbar}\right)^2 +   \mathcal{O}\left[\left(\frac{mLa}{\Delta E}\right)^2\right] \nonumber\,  .
\end{eqnarray}
This is related to the degree of coherence (\ref{eq: deg of coh small acc}), which was calculated as
\begin{equation}
f = \left|\cos^{N-1}\left(\frac{V t}{\hbar} \right)\right| = 1 - \frac{\langle\Delta \omega^2\rangle}{2} t^2 + \mathcal{O}\left[\left(\frac{Vt}{\hbar}\right)^2\right]
\end{equation}
using the density matrix approach in Sec.~\ref{SubSec: Impact of acceleration}.
By that, $1/\sqrt{\langle\Delta \omega^2 \rangle}$ determines the decoherence time scale \cite{pitaevskii2016bose}, which in our case is $\hbar/(V\sqrt{N-1})$.}

\changesOK{
So far, we have discussed the problem of decoherence 
in terms of the pure state frequency $\omega$.
In most of the cases, however, the decoherence is described as a fluctuation of the phase difference $\varphi = \varphi_\textrm{R} - \varphi_\textrm{L}$ between the right and left wells \cite{PhysRevA.55.4318, PhysRevLett.87.180402}. 
Both can be linked,
comparing the expressions (\ref{eq: psi pure}) and (\ref{eqn:order_parameter}) for the pure state in the energy eigenbasis and the left-right basis, respectively. 
From this equality, we obtain
\begin{eqnarray}
    \varphi(\omega) &=& \arctan \left[\frac{\cot(\omega t/2)}{\cos(2\xi)}\right] \\
&=& -\arctan \left[\frac{\Delta E \cot(\omega t/2)}{3maL}\right] + \mathcal{O} \left[\left(\frac{mLa}{\Delta E}\right)^2\right]\, \nonumber\,, \label{eq: wells phase}
\end{eqnarray}
determining the evolution of the relative phase in a pure state case as a function of $\omega$. In the case of vanishing acceleration $a\to0$, the phase becomes $\varphi(\omega)\to -\frac{\pi}{2} \textrm{sign}[\sin\left(\omega t\right)]$. As for the population imbalance (\ref{eqn:pop_imbalance}), the expectation value of $\varphi(\omega)$ and its fluctuations can be obtained by weighted integration together with $P(\omega)$.}





\section{Conclusions}
\label{Sec: Conclusions}

In the present study, we investigated the dynamics of a Bose gas with weak collisional interactions, confined in a double-well potential. We built up our theoretical model from the two-state description of a single particle, which allowed us to construct the basis of many-particle energy eigenstates for the Bose gas. The latter were used to formulate the density matrix, whose evolution describes the dynamics of the double-well system. 
The considered energy-preserving two-particle collisions
give rise to a small perturbation in the equation of motion for the density matrix \cite{blum2012density}.

Having found the evolution of the full $N$-particle density matrix, we derived the population imbalance and the degree of coherence to characterize the Josephson dynamics.
We quantitatively described how the collisions cause the system to decohere and the Josephson oscillations to die off.
These prominent features of collisional decoherence are illustrated \changesOK{with a calculation example}, considering recent experiments with $^{39}$K atoms \cite{masi2021multimode,fattori2008atom}. 

We further analyzed the influence of an external acceleration on the Josephson dynamics and decoherence.
In particular, we studied the effect of small accelerations $a \lesssim 10^{-4} a_\Earth$, which have been previously accessed using BEC accelerometers \cite{masi2021multimode, gersemann2020differential, PhysRevLett.117.203003}.
We found that the dominant effect is the shift of the period of Josephson oscillations, while the collisional decoherence does not notably depend on the acceleration. 
Hence, by measuring the shift of the oscillation period, one can deduce the acceleration applied alongside the Josephson junction. 
This idea of a BEC double-well accelerometer previously was investigated also in Refs.~\cite{roy2025assessing, sacchetti2016accelerated}
assuming a pure state, described by a fully coherent many-particle wave function.
For small accelerations, their numerical simulations show a 
linear change of the oscillation period with acceleration. 
This is in agreement with our analytical results, which \changesOK{furthermore} provide the 
relation between oscillation period, interaction strength, and acceleration for a Bose gas in a given double-well geometry. In the limit of vanishing decoherence, our findings agree with the predictions of the pure state model of standard Josephson equations \cite{Smerzi1997, raghavan1999coherent}.

\changesOK{We have shown how our results, which we obtained from density matrix $\hat{\rho}$ of the many-particle ensemble, can be reformulated in terms of phase fluctuations.
By the established link between the two approaches, we put our results in the broader context of previous research on the decoherence in the Josephson junction \cite{PhysRevA.55.4318, PhysRevLett.87.180402, pitaevskii2016bose,castin1997relative, RevModPhys.90.035005}.}
\clearpage

\changesOK{Even though, a more detailed numerical description may be required, when realistic experiments are considered, our analytical results serve as a benchmark point for numerical methods.}
While our results are formulated in general analytical expressions and can be applied to a wide range of physical scenarios, 
they are obtained under the assumption of 
weak interparticle interactions and small accelerations. It would be an intriguing perspective to investigate the regimes of strong acceleration ($a \sim 10^{-3} a_\Earth$) and strong interaction ($a_s \sim 10^2a_0$), where the Josephson dynamics is expected to differ significantly, as implied by recent BEC experiments \cite{masi2021multimode, gersemann2020differential, PhysRevLett.118.230403}. 
\changesOK{ 
Moreover, to extend the formalism towards open systems would allow to take into account heating due to external influences, that may be a dominant channel of decoherence in a weakly interacting system.}







\acknowledgements{
We thank A. Yakimenko and A. Surzhykov for helpful suggestions and Y. Bidasyuk for fruitful discussions about the pure state model.
We are also grateful to
N. Gaaloul, K. Frye-Arndt and B. Rhyno for inspiring discussions on BECs in box traps.
Funded by the Deutsche Forschungsgemeinschaft (DFG, German Research Foundation) under Germany’s Excellence Strategy—EXC 2123 QuantumFrontiers—390837967.}

\vspace{5 mm}

\appendix

\section{Potential geometry and single-particle states}
\label{Appendix: Potential geometry and single-particle states}

The single particle Hamiltonian reads
\begin{equation}
\hat h_0 (\mathbf{r}) = -\frac{\hbar^2}{2m}\left(\p_x^2 +\p_y^2 + \p_z^2\right) + U_x(x) + U_y(y) + U_z(z)\,, \label{eqn:one_particle_Hamiltonian}
\end{equation}
where the expressions for the potential components $U_x (x)$, $U_y (y)$ and $U_z (z)$ are illustrated by Fig.~\ref{Fig: trap geometry}.

For $U_z(z)$ we have
\begin{equation}
U_z(z)= \left\{\begin{array}{ll}
0 & \quad z \in [-\infty, -2L] \\
       - U_0 & \quad z \in [-2L, -L] \\
         0 & \quad z \in [-L, L] \\
       -U_0  & \quad z \in [L, 2L]\\
      0 & \quad z \in [2L, +\infty] \,\,,
       \end{array}\right.\, 
       \label{Eq: double bin Vx}
\end{equation}
\vspace{0.05 cm}
which is a double-well potential, while in the other two dimensions we have trivial confinement
\begin{equation}
U_x(x)= \left\{\begin{array}{ll}
0& \quad x \in [-\infty, -L/2] \\
        -U_0 & \quad x \in [-L/2, L/2] \\
       0 & \quad x \in [L/2, +\infty] \,\,,
       \end{array}\right.\,
       \label{Eq: double bin Vy}
\end{equation}
\begin{equation}
U_y(y)= \left\{\begin{array}{ll}
0 & \quad y \in [-\infty, -L/2] \\
        -U_0 & \quad y \in [-L/2, L/2] \\
       0 & \quad y \in [L/2, +\infty] \,\,.
       \end{array}\right.\,
       \label{Eq: double bin Vz}
\end{equation}

Thus, we end up with the two cubic potential traps, separated by the space of width $2L$. For such a choice of Hamiltonian the single-particle state wave function factorizes $\phi(x,y,z) = \phi_x(x)\phi_y(y)\phi_z(z)$ and has the energy $E = E_x + E_y + E_z$. For convenience, we introduce dimensionless units of distance $x = L \tilde{x}$, $y = L \tilde{y}$ and $z = L \tilde{z}$, dimensionless wave function components $\tilde{\phi}_{x,y,z} = \sqrt{L}\phi_{x,y,z}$ (so that it is normalized as $\int |\tilde{\phi}_{x}|^2 d \tilde{x} = 1$). The dimensionless energy then reads $\tilde{E}_{x,y,z} = 2mL^2/\hbar^2 \times E_{x,y,z}$ and dimensionless potential is $\tilde{U}_0 = 2mL^2/\hbar^2 \times U_0$.
This gives us the dimensionless form of Schrodinger equation
\begin{equation}
\tilde{E}_{x,y,z}\tilde{\phi}_{x,y,z} = \left(-\p_{\tilde{x},\tilde{y},\tilde{z}}^2 + \tilde{U}_{x,y,z}\right)\tilde{\phi}_{x,y,z} \, ,
\label{app eq: Schrodinger equation}
\end{equation}
which for bound states gives negative energies $\tilde{E}_{x,y,z} = - \tilde{k}^2_{x,y,z} < 0$ ($\tilde{k} = Lk$ is the dimensionless wavevector). Let's for brevity denote $\tilde{\phi}_{x,y} = \tilde{\phi}_\perp$ and $\tilde{E}_{x,y} = \tilde{E}_\perp$. 

In transverse direction Eq.~(\ref{app eq: Schrodinger equation}) yields solution 
\begin{equation*}
\tilde{\phi}_\perp(x)= \left\{\begin{array}{ll}
       C e^{\tilde{k}_x \tilde{x}} & \quad x \in [-\infty, -L/2] \\
        D\cos \left(\sqrt{\tilde{U}_0 - \tilde{k}_x^2} \tilde{x} \right) & \quad x \in [-L/2, L/2] \\
       C e^{-\tilde{k}_x \tilde{x}} & \quad x \in [L/2, +\infty]\,\,.
       \end{array}\right.\,
\end{equation*}
Here the dimensionless momentum $\tilde{k}_x$ can be found numerically as a solution of transcendental equation \cite{binney2013physics}, while the coefficients $C$ and $D$ are found from the continuity relations and normalization condition. In $z-$direction solution of Eq.~(\ref{app eq: Schrodinger equation}) reads
\begin{widetext}
\begin{equation*}
\tilde{\phi}_z(z)= \left\{\begin{array}{ll}
       A_1 e^{\tilde{k}_z \tilde{z}} & \quad z \in [-\infty, -2L] \\
      A_2\cos \left(\sqrt{\tilde{U}_0 - \tilde{k}_z^2} \tilde{z} \right) + B_2 \sin \left(\sqrt{\tilde{U}_0 - \tilde{k}_z^2} \tilde{z} \right) & \quad z \in [-2L, -L] \\
        A_3 e^{\tilde{k}_z \tilde{z}} + B_3 e^{-\tilde{k}_z \tilde{z}} & \quad z \in [-L, L] \\
        A_4 \cos \left(\sqrt{\tilde{U}_0 - \tilde{k}_z^2} \tilde{z}\right) + B_4 \sin \left(\sqrt{\tilde{U}_0 - \tilde{k}_z^2} \tilde{z} \right) & \quad z \in [L, 2L] \\
       B_5 e^{-\tilde{k}_z \tilde{z}} & \quad z \in [2L, +\infty]\,\,,
       \end{array}\right.\,
\end{equation*}
\end{widetext}
which gives the two ground and excited states $\tilde{\phi}^z_{0/1}$. The corresponding values of $\tilde{k}_z$ are derived numerically as solutions of the transcendental equation, while all the coefficients $A_{1/2/3/4}$ and $B_{2/3/4/5}$ are defined by the continuity relations together with normalization condition.

\begin{figure}[htp]
    \centering  \includegraphics[scale = 0.4]{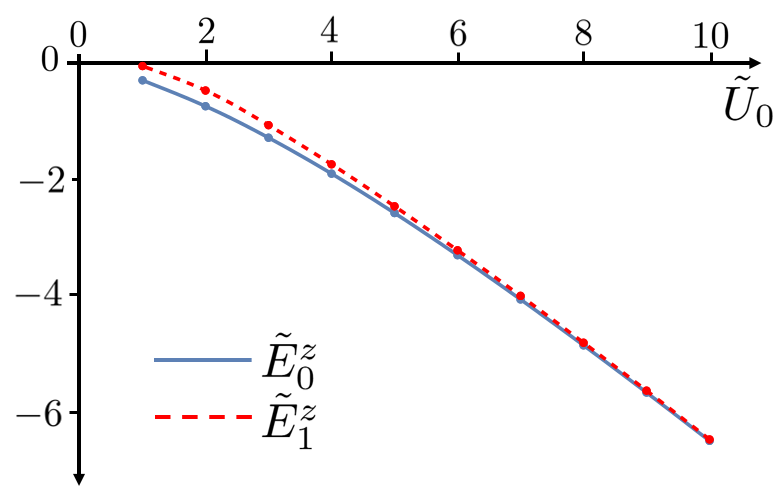}
    \caption{Energies $\tilde{E}_0^z$ and $\tilde{E}_1^z$ depending on the depth of a double-well $\tilde{U}_0$}
    \label{Fig: Energy levels}
\end{figure}

Staying within the two-state model we will consider the choice $\tilde{U_0} = 5$ of the dimensionless potential depth. In this case in longitudinal $z$ direction two bound states exist with the dimensionless wavevectors
$\tilde{k}_0 = 1.6034$, $\tilde{k}_1 = 1.5664$ and energies $\tilde{E}^z_0 = -2.5709$, $\tilde{E}^z_1 = -2.4536$. In transverse $x-y$ directions only one bound state exists with the 
dimensionless wavevector $\tilde{k}^\perp = 1.5857$ and energy $\tilde{E}^\perp = -2.514$. Thus, for the specific trap geometry we have two states with the energy gap $\Delta \tilde{E} = 0.117$ between them. The energies $\tilde{E}_{0/1}^z$ for other choices of $\tilde{U}_0$ are illustrated in Fig.~\ref{Fig: Energy levels}.

\section{Beyond two state model}
\label{Appendix: Beyond two state model}

One can generalize the two-state model to a many-state case. Assume, we have $k>2$ single-particle energy eigenstates $\phi_i$ in  a double-well trap. In the non-interacting limit this gas is described by the Hamiltonian
\begin{equation}
    \hat{H}_0 = \sum_{i=0}^k E_i \hat{a}_i^\dagger \hat{a}_i\, ,
\end{equation}
where $E_i$ are the eigenenergies corresponding to the $\phi_i$ states. The state of the system is then described by
\begin{eqnarray*}
    \hat{\rho} &=& \sum_{\{N_i\},\{N_i'\}} \rho_{\{N_i\},\{N_i'\}}|\{N_i\} \rangle \langle \{N_i'\} |  \\
    |\{N_i\}\rangle &=& \frac{1}{\sqrt{N_0! N_1! ... N_k!}} \left(\hat{a}_0^\dagger \right)^{N_0} ... \left(\hat{a}_k^\dagger \right)^{N_k}|0\rangle \, ,
\end{eqnarray*}
where each many-particle state $|\{N_i\}\rangle$ is characterized by a set of occupations $\{N_i\} = \{N_0,..., N_k\}$ of all energy eigenstates. The interaction operator in this case reads
\begin{eqnarray*}
    \hat{\mathcal{V}}_I &=& \hat{\mathcal{V}} = \frac{1}{2} \sum_{i j k l} V_{i j k l} \hat{a}_i^\dagger \hat{a}_j^\dagger \hat{a}_k \hat{a}_l 
    =  \frac{1}{2} \sum_{i} V_{i i i i} \hat{a}_i^\dagger \hat{a}_i^\dagger \hat{a}_i \hat{a}_i \\ &+& \frac{1}{2} \sum_{i\neq j} \left[V_{i j i j} + V_{i j j i} + V_{j i i j} + V_{j i j i}\right] \hat{a}_i^\dagger \hat{a}_j^\dagger \hat{a}_i \hat{a}_j \, ,
\end{eqnarray*}
where we assumed that the energy level scheme $E_i$ is such that the energy conservation is satisfied only if the two outgoing bosons have the same energies as the ingoing bosons or if they 'exchange' their energies (analogously to Fig.~\ref{Fig: Collision diagrams}). 

Solving Liouville equation~(\ref{eq: Liouville equation}) we find the solution $\rho_{\{N_i\},\{N_i'\}} = \rho^0_{\{N_i\},\{N_i'\}} \exp[i \omega_{\{N_i\},\{N_i'\}}t]$, where $\omega_{\{N_i\},\{N_i'\}} = (\mathcal{E}_{\{N_i'\}} - \mathcal{E}_{\{N_i\}} - \mathcal{V}_{\{N_i\},\{N_i'\}})/\hbar$ reads
\begin{eqnarray}
    \mathcal{E}_b - \mathcal{E}_a &=& \sum_i (N_i' - N_i) E_i \\
    \mathcal{V}_{ab} &=& \frac{1}{2} \sum_i V_{iiii} \left(N_i (N_i - 1) - N_i' (N_i' - 1)\right) \nonumber\\
    &+& \sum_{i \neq j} V_{ijij} \left(N_i N_j - N_i' N_j'\right)
\end{eqnarray}

In analogy with the two-state case we introduce the optimal left/right well basis, which allows us to define population imbalance in a symmetric double-well
\begin{eqnarray*}
    Z (t) &=& \frac{4}{N}\sum_{\kappa, \eta} \langle \phi_{2(\kappa-1)}|\phi_{2\eta-1}\rangle_L \\
    &\times& \textrm{Re} \left[ \alpha_{2(\kappa-1), 2\eta-1}^0 e^{i \omega_{2(\kappa-1), 2\eta-1} t}\right],
    \label{eq: left well occupation for k states}
\end{eqnarray*}
in the non-interacting case. Here $\langle \phi_{i}|\phi_{j}\rangle_L = \int dx_{-\infty}^{+\infty} dy_{\infty}^{+\infty} dz_{-\infty}^0 \phi_{i}(\mathbf{r}) \phi_{j}(\mathbf{r})$ is the left well overlap of the two single-particle states. This result illustrates that the role of the mixing $\rho^0_{i,j}$ between an even $i=2(\kappa-1)$ and an odd $j = 2\eta-1$ states is weighted by the left overlap between them.

\section{Matrix elements of collisions}
\label{Appendix: Collisional matrix elements}

\begin{figure}[htp]
    \centering  \includegraphics[scale = 0.48]{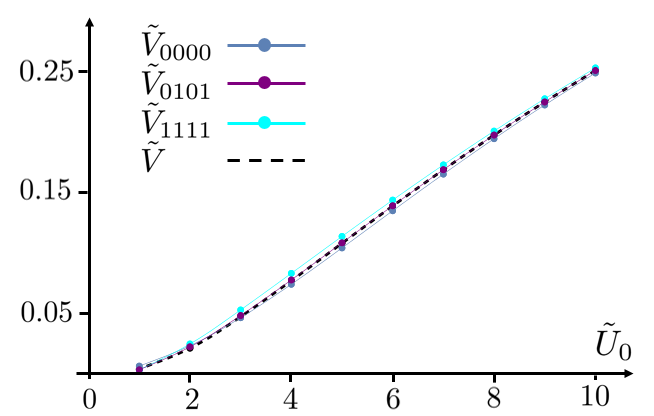}
    \caption{Dimensionless matrix elements $\tilde{V}_{ijrl}$ depending on the depth $\tilde{U}_0$ of the potential. The approximate average value $\tilde{V}$ is calculated using (\ref{eq: matrix elemnt V})}
    \label{Fig: V matrix elements}
\end{figure}

For the wave functions $\phi_{0/1}(x,y,z)$ discussed in Appendix~\ref{Appendix: Potential geometry and single-particle states}, we consider a matrix element of two-particle collision
\begin{eqnarray*}
    V_{ijrl} &=& \int d \mathbf{r}_1 d \mathbf{r}_2 \phi^\star_i(\mathbf{r}_1) \phi^\star_j(\mathbf{r}_2) V(\mathbf{r}_1 - \mathbf{r}_2) \phi_r (\mathbf{r}_2) \phi_l (\mathbf{r}_1) \nonumber \\ &=& g \int d \mathbf{r} \phi^\star_i(\mathbf{r}) \phi^\star_j(\mathbf{r}) \phi_r (\mathbf{r}) \phi_l (\mathbf{r})\, 
\end{eqnarray*}
with two ingoing $\phi_{i/j}^\star$ and two outgoing $\phi_{r/l}$ bosons. For convenience we rewrite it in dimensionless units as
\begin{equation*}
    V_{ijrl} = gL^3 \left(\frac{1}{\sqrt{L}}\right)^{12}\int d\tilde{x} d\tilde{y} d\tilde{z} \tilde{\phi}_i^\star \tilde{\phi}_j^\star \tilde{\phi}_r \tilde{\phi}_l  = \frac{g}{L^3} \tilde{V}_{ijrl}.
\end{equation*}
For a deep double-well, i.e. large $\tilde{U}_0$, in the absence of acceleration we have $\tilde{V}_{0000} = \tilde{V}_{1111} = \tilde{V}_{0101} = \tilde{V}_{1010} = \tilde{V}$. 
In the currently suggested geometry (see Sec.~\ref{Sec: Two-state model for 3D cubic traps}) the matrix elements depend on the finite height of potential well $\tilde{U}_0$ and can be calculated numerically or using approximate expression
\begin{equation}
    \tilde V = \frac{1}{2}\left(\frac{\sqrt{3\tilde{E}}\left(4 \tilde{U}_0 + \tilde{U}_0 \sqrt{3\tilde{E}} + 2 \tilde{E}/3\right)}{2  \tilde{U}_0 (2 \sqrt{3} + \sqrt{\tilde{E}})^2}\right)^3\, ,
    \label{eq: matrix elemnt V}
\end{equation}
where the energy $\tilde{E} = |\tilde{E}_0 + \tilde{E}_1|/2$ has to be found numerically. This dependency on $\tilde{U}_0$ for nonzero $\tilde{V}_{ijrl}$ is illustrated in Fig.~\ref{Fig: V matrix elements}.

When the potential depth $\tilde{U}_0$ increases, the wave functions $\tilde{\phi}_{0/1}$ are more localized and have more pronounce peaks within the wells of the double-well. This leads to the increase of the matrix elements $\tilde{V}_{ijrl}$, as shown in Fig.~\ref{Fig: V matrix elements}. The approximate equality $\tilde{V}_{0000} = \tilde{V}_{1111} = \tilde{V}_{0101}$ holds true with accuracy $1.6 \%$ for a deep double-well $\tilde{U}_0 = 10$ and $6 \%$ for a shallow double-well with $\tilde{U}_0 =2$. For the particular value $\tilde{U}_0 = 5$, that we use to illustrate our results, we have $\tilde{V}_{0000} = 0.1051$, $\tilde{V}_{1111} = 0.1148$ and $\tilde{V}_{0101} = 0.1094$, in our calculations we use an approximate same value of $\tilde{V} = 0.1$ for all those.

\section{Pure state Josephson effect}
\label{Appendix: Comparison with the pure state approach}

\begin{figure}[htp]
    \centering  \includegraphics[scale = 0.4]{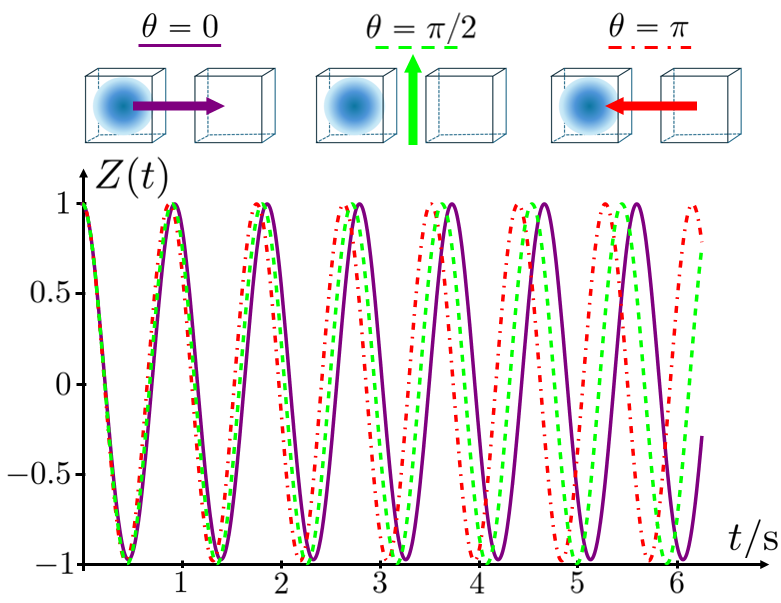}
    \caption{Numerical solution of Josephson equations (\ref{eq: Jeap1}-\ref{eq: Jeap2}) for $\Delta E/\hbar = 7.06\, \textrm{s}^{-1}$ and $\Lambda = -0.605$. The purple, green and red lines correspond to the cases with acceleration $\mathbf{a}$ directed along $\theta = 0$, $\theta = \pi/2$ and $\theta = \pi$ with $\Xi = -0.094$, $\Xi = 0$ and $\Xi = 0.094$ respectively. This case corresponds to a $^{39}$K setup discussed in Sec.~\ref{subsec: Josephson dynamics without external acceleration} with interaction strength $g = 15g_0$.}
    \label{Fig: num sol of JEs}
\end{figure}

In this Appendix we discuss the standard two-state model of Josephson effect in more detail. We plug the ansatz $\Psi = a_L(t)\psi_L(\mathbf{r}) + a_R(t)\psi_R(\mathbf{r})$ into Eq.~(\ref{eq: GP}), and integrate the resulting equation with $\int d\mathbf{r} \psi_L \times$ and  $\int d\mathbf{r} \psi_R \times$. Then, neglecting all integrals that contain the overlap of $\psi_{L}$ and $\psi_R$ states we obtain 
\begin{eqnarray}
    i\hbar \dot{a_L} &=& \left[\epsilon_L + vN_L\right]a_L - K a_R \label{eq: C1}\\
    i\hbar \dot{a_R} &=& \left[\epsilon_R + vN_R\right]a_R - K a_L  \, ,\label{eq: C2}
\end{eqnarray}
where the parameters read
\begin{eqnarray*}
    \epsilon_L &=& E_0^a \cos^2 \xi + E_1^a \sin^2 \xi\\
    \epsilon_R &=& E_0^a \sin^2 \xi + E_1^a \cos^2 \xi \\
    v &=& g\int d\mathbf{r}|\psi_L|^4 = g\int d\mathbf{r}|\psi_R|^4\\
    K &=& (E_0^a - E_1^a)\sin \xi \cos \xi
\end{eqnarray*}
and we used the fact that $\psi_{L/R}$ are related to the energy eigenstates $\psi_{0/1}$ via transformation (\ref{eq:left-right states}).
Defining $a_{L/R}(t) = \sqrt{N_{L/R}(t)}e^{i\varphi_{L/R}(t)}$ and introducing population imbalance $Z(t)=(N_L(t) - N_R(t))/N$ and phase difference $\varphi(t) = \varphi_R(t) - \varphi_L(t)$, we rewrite equations (\ref{eq: C1}) and (\ref{eq: C2}) as
\begin{eqnarray}
    \frac{\hbar}{2K} \dot{Z} &=& -\sqrt{1 - Z^2} \sin \varphi \label{eq: Jeap1}\\
    \frac{\hbar}{2K} \dot{\varphi} &=& \Xi + \Lambda Z + \frac{Z}{\sqrt{1 - Z^2}}\cos \varphi \, . \label{eq: Jeap2}
\end{eqnarray}
The latter equations are well known as standard Josephson equations \cite{Smerzi1997} and contain the two parameters $\Xi = (\epsilon_L - \epsilon_R)/(2K)$ and $\Lambda = vN/(2K)$. For small accelerations we keep only linear order corrections in $maL/\Delta E$, which allows us to simplify
\begin{eqnarray*}
    \epsilon_L - \epsilon_R &=& (E_0^a - E_1^a)\cos(2\xi) \approx 3aLm\cos \theta\\
    K &=& \frac{1}{2}(E_0^a - E_1^a)\sin(2\xi) \approx -\frac{\Delta E}{2}\, ,
\end{eqnarray*}
giving $\Xi \approx -\frac{3aLm}{\Delta E} \cos \theta$. For the $\Lambda$ parameter, characterizing the strength of interparticle interactions we find $\Lambda \approx VN/K = -2VN/\Delta E$. This gives us Eqs.~(\ref{eq: main Jos eqs}), whose numerical solution is illustrated in Fig.~\ref{Fig: num sol of JEs}. 
In the Figure we see that $Z(t)$ exhibits harmonic oscillations with the frequency of these oscillations being increased or decreased, depending on the direction of $\mathbf{a}$. This frequency shift with respect to its unperturbed value $\Delta E/\hbar$ (green dashed line in Fig.~\ref{Fig: num sol of JEs}) can be well approximated by $\pm \Delta E \Lambda \Xi/(2\hbar)$.

\section{\changesOK{Density matrix and phase fluctuations}}
\label{Appendix: Density matrix and phase fluctuation}

Following the discussion in Sec.~\ref{Sec: Decoherence}, we first find the number of bosons occupying ground and excited energy states $\alpha_{11} = N \sin^2 \xi$ and $\alpha_{00} = N \cos^2 \xi$ as well as the nondiagonal element of $\hat{\rho}_e$
\begin{eqnarray*}
    \alpha_{01}(t) &=& N!(\cos \xi)^{2N} e^{i \Delta E^a t/\hbar} \times\\
    & &\sum_{N_1 = 0}^{N-1} \frac{(\tan \xi)^{2N_1 + 1}}{N_1 ! (N - N_1 - 1)!} e^{i V^a(N - 2N_1 - 1)t/\hbar} \\
    &=& \frac{N}{2} \sin(2 \xi) e^{i \Delta E^a t/\hbar} e^{i(N-1)V_a t/\hbar} \times \\
    & &(\cos^2 \xi + e^{-2iV^a t/\hbar} \sin^2 \xi)^{N-1} \, .
\end{eqnarray*}
Using Eq.~(\ref{eq: alph01phfluct}) we obtain
\begin{eqnarray*}
    \frac{\alpha_{01}(t)}{\sqrt{\alpha_{00} \alpha_{11}}} &=& \frac{2}{N \sin(2\xi)} \alpha_{01}(t) \\
    &=& e^{i \Delta E^a t/\hbar} e^{i(N-1)V_a t/\hbar}  (\cos^2 \xi + e^{-2iV^a t/\hbar} \sin^2 \xi)^{N-1} \, ,
\end{eqnarray*}
and applying inverse Fourier transform we find
\begin{eqnarray*}
P(\omega) &=& \frac{1}{2\pi}\int_{-\infty}^{+\infty} dt e^{-i \omega t} \frac{\alpha_{01}(t)}{\sqrt{\alpha_{00} \alpha_{11}}} 
= \hbar \left(\sin \xi\right)^{2(N-1)}  \\
&\times& \sum_{k=0}^{N-1} C^k_{N-1}\cot^{2k}\xi \delta(\hbar \omega - \Delta E^a - V^a (2k-N+1))\, .
\end{eqnarray*}
\changesOK{The average frequency of this distribution reads 
\begin{eqnarray}
    \omega_\textrm{av} &=& \int_{-\infty}^{+\infty} d\omega P(\omega) \omega = \frac{\Delta E^a}{\hbar} + \frac{V^a}{\hbar}(N - 1)\cos(2 \xi) \nonumber \\
    = \frac{\Delta E}{\hbar} &-& \frac{3Lma \cos \theta}{\Delta E} \frac{V}{\hbar} (N-1) +   \mathcal{O}\left[\left(\frac{mLa}{\Delta E}\right)^2\right]\, \label{eq: omega av} ,
\end{eqnarray}
where the latter formula is an approximate expression in the regime $a \ll \Delta E/(Lm)$.}

\changesOK{The} fluctuation of the $P(\omega)$ distribution \changesOK{can be characterized by its variance
\begin{eqnarray*}
 \langle \Delta \omega^2 \rangle &=& \int_{-\infty}^{+ \infty} d\omega P(\omega) \left(\omega - \omega_\textrm{av}\right)^2 \\ &=& (N-1)\left(\frac{V^a}{\hbar}\right)^2 
 \sin^2 \left(2\xi\right) \\ &= &  (N-1)\left(\frac{V}{\hbar}\right)^2 +   \mathcal{O}\left[\left(\frac{mLa}{\Delta E}\right)^2\right]\, ,
\end{eqnarray*}
where the latter expression is the approximation in the regime of small accelerations.}

Averaging the pure state population imbalance (i.e., in the absence of interactions, $V^a = 0$ and $\omega = \Delta E^a/\hbar$)
\begin{equation*}
    Z_\textrm{pure}(\omega, t) = \cos^2(2 \xi) + \sin^2 (2\xi) \cos \left(\omega t\right)
\end{equation*}
with the distribution $P(\omega)$ we find $\langle Z(t) \rangle = \int d\omega P(\omega) Z_\textrm{pure}(\omega, t)$, which coincides with our result (\ref{eqn:Zoftwithchi}).
\changesOK{In the regime $a \ll \Delta E/(Lm)$ we can directly see that $Z_\textrm{pure}(\omega_\textrm{av}, t) \approx \cos(\omega_\textrm{av}t)$, with the average frequency given by (\ref{eq: omega av}), recovers our result (\ref{eq:Zfor smallacc}) up to the decoherence factor $\cos^{N-1}(Vt/\hbar)$.}

\bibliography{main}

\end{document}